\documentstyle[12pt,aasms4]{article}

\lefthead{Lipari et al.}

\righthead{Multiple Merger, Galactic-Wind and Inflow in NGC 3256}

\begin{document}

\begin{center}

{\bf LUMINOUS INFRARED GALAXIES. III. MULTIPLE MERGER,} \\

\vspace{5mm}

{\bf  EXTENDED MASSIVE STAR FORMATION, GALACTIC-WIND }\\

\vspace{5mm}

{\bf AND NUCLEAR-INFLOW IN NGC~3256 \altaffilmark{1} }\\

\end{center}

\vspace{25mm}

\altaffiltext{1}{Based on observations obtained at the 
              Hubble Space Telescope (HST-WFPC2 \& NICMOS) satellite, 
              International Ultraviolet Explorer (IUE) satellite, 
              European Southern Observatory (ESO-NTT) Chile,
              Cerro Tololo Interamerican Observatory (CTIO) Chile, 
              Complejo Astron\'omico el Leoncito (CASLEO) Argentina, 
              Estaci\'on Astrof\'{\i}sica de Bosque Alegre (BALEGRE) Argentina.}

\author{S. L\'{\i}pari \altaffilmark{2,3},
        R. D\'{\i}az \altaffilmark{2,3}, 
        Y. Taniguchi \altaffilmark{4}, 
        R. Terlevich \altaffilmark{5}, 
        H. Dottori \altaffilmark{3,6}, and 
        G. Carranza \altaffilmark{2}}

\altaffiltext{2}{Cordoba Observatory and CONICET, Laprida 854, 5000 Cordoba,
Argentina (lipari@mail.oac.uncor.edu).} 

\altaffiltext{3}{Visiting astronomer at BALEGRE, CASLEO, CTIO and ESO
Observatories.}

\altaffiltext{4}{Astronomical Institute, Tohoku University, Aoba, Sendai 
980-8578, Japan.}

\altaffiltext{5}{Institute of Astronomy, Madingley Road, Cambridge CB3 0HA,
United Kingdom.}

\altaffiltext{6}{Instituto de Fisica, Univ. Fed. Rio Grande do Sul, CP 15051,
Porto Alegre, Brazil.}

\begin{abstract}
We report detailed evidence for: multiple merger, extended massive star 
formation, galactic--wind and circular/non-circular motions in the luminous 
infrared galaxy NGC 3256. Based on observation of high resolution imaging
(obtained at HST and ESO--NTT), and extensive kinematical/spectroscopic
data (more than 1000 spectra, collected at Bosque Alegre, CASLEO, CTIO
and IUE Observatories).

We find in a detailed {\it morphological} study (at $\sim$15 pc 
resolution) that the extended massive star formation process, detected 
previously in NGC 3256 shows:
(i) extended triple asymmetrical spiral arms structure (r $\sim$ 5 kpc); and
(ii) the spiral arms emanate from three different nuclei.
The main optical nucleus shows a small spiral--disk (r $\sim$ 500 pc) which is 
a continuation of the external one and reach the very nucleus. And this very 
nucleus shows blue elongate structure (63 pc $\times$ 30 pc), and
luminous blue star cluster properties. We discuss this complex morphology, in 
the framework of an extended massive star formation driven by multiple merger 
process (Hernquist et al., Taniguchi et al.'s models).   

We study the {\it kinematics} of this system and
present a detailed H$\alpha$ velocity field for
the central region (40$''\times$40$''$; r$_{max} \sim$30$'' \sim$5 kpc); 
with a spatial resolution of 1$''$, and errors of $\pm$15 km~s$^{-1}$. 
The color and isovelocity maps show mainly: 
(i) a clear kinematical center of circular motion with ``spider" shape and 
located between the main optical nucleus and the close (5$''$) mid-IR 
knot/nucleus; (ii) non--circular motions in the external parts. 
In the main optical nucleus we found a clear {\it ``outflow component" 
associated to galactic--winds} and a  {\it ``inflow radial motion"} (in 
the spiral-disk nuclear structure, r $\sim$700 pc). 
In addition, we detected the outflow component in the central and external
regions (r $\leq$ 5-6 kpc), with a very wide opening angle $\theta$ =
140$^{\circ}$. We found that the mean value of the inflow region
(at PA $\sim$80$^{\circ}$) is practically perpendicular to the axis of the
bipolar outflow (at PA $\sim$160$^{\circ}$).
Our optical spectroscopic data cube show that the
three asymmetrical spiral arms have no kinematical relation to each other. 

We analyze in detail the {\it physical conditions} in the giant H {\sc ii} 
regions located in the asymmetric spiral arms, the two main optical knots/
nuclei, and the outflow component (using long slit spectroscopy,
 plus standard models of fotoionization, shocks and starburst). 
We present four detailed emission line ratios (N {\sc ii}/H$\alpha$,
S {\sc ii}/H$\alpha$, S {\sc ii}/S {\sc ii}) and FWHM(H$\alpha$) maps
for the central region (30$''\times$30$''$; r$_{max} \sim$22$'' \sim$4 kpc), 
with a spatial resolution of 1$''$. We found that
the massive star formation is extended from the very nucleus to the end of 
the tidal tails: i.e., from r $\sim$15 pc to $\sim$40 kpc. In particular, 
in the central region (r $\sim$ 5--6 kpc) we detected that the nuclear starburst
and the extended giant H{\sc ii} regions (in the spiral arms) have very similar 
properties, i.e., high metallicity and low ionization spectra, with: 
T$_{eff}$ = 35000 $^{\circ}$K, solar abundance, a range of T$_{e}$
$\sim$6000--7000 $^{\circ}$K and N$_{e}$ $\sim$100--1000 cm$^{-3}$.
These results
are in excellent agreement with the global IR emission lines studies made
by Joseph and collaborators. In addition, the nuclear and extended outflow
shows properties typical of galactic--wind/shocks, associated to the
nuclear starburst. And, we suggest that the interaction 
between dynamical effects, the galactic--wind (outflow), low-energy cosmic
rays, and the molecular+ionized gas (probably
in the inflow phase) could be the possible mechanism that generate the 
{\it ``similar extended properties in the massive star formation, 
at scale of 5-6 kpc!"}.

In conclusion, these results suggest that NGC 3256
is the product of a multiple merger, which generated an extended massive star
formation process with an associated galactic--wind plus an inflow
(mainly, in the nuclear region).
Finally, we have also studied the presence of the close merger/interacting
systems NGC 3256C (at $\sim$150 kpc, and $\Delta$V $\sim$ 200 km s$^{-1}$),
and NGC 3256A, plus the possible association between the NGC 3256 and 3263
groups of galaxies. Furthermore, we analyze for NGC 3256
the possible evolution  from luminous IR galaxy to  QSOs, elliptical, cD, 
or radio galaxy (Toomre, Schweizer, Joseph et al., Sanders et al., Terlevich
et al.'s models), where {\it the powerful galactic--wind  and the
relation between mergers and extreme dusty-starburst}
play a main role in this evolutive process (Rieke et al., Joseph et al.,
Heckman et al., Lipari et al.'s hypothesis), and probably in the formation
and evolution of galaxies.

\end{abstract}

\keywords{ 
galaxies: evolution -- galaxies: individual (NGC~3256, NGC~3256C, NGC~3256A, 
NGC~3256B, NGC~3263, NGC~3262) -- galaxies: interactions -- 
galaxies: kinematics -- galaxies: starburst -- infrared: galaxies -- 
quasar: general  
} 

\newpage

\section{INTRODUCTION}

One important current issue in astrophysics and cosmology (specially at high 
redshift) is the study of {\it star formation in merger process} and their
relation to the {\it evolution and formation of galaxies} (see for references
Djorgovski 1994; White 1994; Weil \& Hernquist 1996; Scott 1998).
Furthermore, this issue is also related to the detection of highly 
energetic activity in the nuclei of galaxies (see Toomre \& Toomre 1972;
Toomre 1977; Rees 1984; Joseph \& Wright 1985; Sanders et al. 1988a,b;
Norman \& Scoville 1988; Taniguchi, Ikeuchi \& Shioya 1999). On the other
hand, detailed studies of nearby mergers give important clues, in order 
to understand properly high redshift mergers and merger-candidates.

Luminous infrared (IR) galaxies (L$_{IR} \geq 10^{11} L_{\odot}$; LIRGs) are 
dusty strong IR emitters (L$_{IR}$/L$_{B} \sim$ 10--300) 
where frequently an enhancement of star formation is taking place 
(see Rieke \& Low 1972, 1975; Rieke et al. 1980, 1985; Weedman et al. 1981;  
Joseph \& Wrigth 1985; Norris et al. 1990; 
Heckman, Armus, \& Miley 1990 -hereafter HAM90-; Hutching \& Neff 1991; 
Condon et al. 1991a,b; Lipari, Terlevich \& Macchetto 1993a; 
Lutz et al. 1996; Downes \& Solomon 1998).
Furthermore, in LIRGs, ``galactic--winds/bubbles" features 
originating in massive stars and/or supernova explosions have been also 
detected (see Ulrich 1972, 1978; Heckman, Armus, \& Miley 1987, HAM90; 
Mc-Carthy, Heckman, \& van Breugel 1987; Bland \& Tully 1988; Taniguchi 
et al. 1988; Fabiano 1988; Colina, Lipari \& Macchetto 1991; Filippenko \& 
Sargent 1992; Phillips 1993; Lipari, Colina, \& Macchetto 1994; 
Veilleux et al. 1994; Lipari, Tsvetanov \& Macchetto 
1997). However, at high IR luminosities there is also an increase of the 
non--thermal nuclear activity (Sanders et al. 1988a; Veilleux et al. 1996,
1999).

On the other hand, imaging surveys of ultraluminous IR galaxies  
(L$_{IR} \geq 10^{12} L_{\odot}$; ULIRGs) show that $\sim$100$\%$ 
are mergers/interacting systems 
(Sanders et al. 1988a,b; Melnick \& Mirabel 1990; Clements et al. 1996;
Joseph \& Wright 1985; Rieke et al. 1985).
In addition, virtually all LIRGs have been shown to be 
extremely rich in interstellar molecular gas (Sanders, Scoville, \& Soifer 
1991) and highly concentrated in their nuclei (Scoville et al. 1991; Okumura 
et al. 1991; Downes \& Solomon 1998). This matter reservoir plays a central 
role in promoting star formation and probably fueling a massive compact 
object (Scoville \& Soifer 1991).

Mergers are mainly luminous IR galaxies where their luminosities overlap with 
most luminous QSOs and Seyfert galaxies; and their optical, IR and radio properties
show mainly starburst and AGN characteristics (see Joseph \& Writh 1985; 
Schweizer 1980, 1982, 1990; Rieke et al. 1985; Heckman, Armus \& Miley 1987; 
Sanders et al. 1988a; Colina et al. 1991; Lipari, 
et al. 1994; Veilleux et al. 1996, 1999; Lutz et al. 1996; Genzel et 
al. 1998). And, there is compelling evidence for 
merger/interaction driven {\it starburst and nuclear activity}; probably by 
depositing large amount of interstellar gas to the central regions (see 
Sanders \& Mirabel 1996; Scoville \& Soifer 1991; Barnes \& Hernquist 1992; 
Mihos \& Hernquist 1996, 1994a,b). Furthermore, there are also strong
evidences for galactic--wind features, from young starburst in luminous IR 
mergers (see \S 4.5).
And, several possible links between mergers, starburst,
IR~galaxies, AGNs, QSOs and evolution of galaxies have been proposed 
(see Weedman 1983; Rees 1984; Norman \& Scoville 1998; Perry 1992; Perry \&
Dyson 1992; Lipari 1994; Taniguchi et al. 1999). 

NGC~3256 is a nearby luminous IR galaxy (z = 0.0094; \S 3.2) which 
has been thought
to be a major merger between two gas rich galaxies (Toomre 1977; Graham et al.
 1984; Joseph \& Writh 1985; Schweizer 1986, 1990; Joseph 1991; Sanders et al. 
1995). This system shows knotty circumnuclear structure (Sersic 1959)
 and two extended tidal tails (de Vaucouleurs \& de Vaucouleurs 1961).
The total IR luminosity of NGC~3256 is L$_{IR[8-1000 \mu m]} = 3.3 \times 
10^{11} L_{\odot}$, consequently the 
bolometric luminosity 
makes this system the most luminous nearby galaxy (for V $\leq$ 3000 
km~s$^{-1}$; Sargent, Sanders, \& Phillips 1989). 

The strong 10$\mu$m IR emission in NGC~3256 is {\it ``very extended"}, with 
most of it originating outside the central kpc, indicating {\it clearly that 
the near--IR emission is powered mainly by a recent starburst}, possibly 
triggered by the merger process (Graham et al. 1984; Joseph \& Wright 1985; 
Joseph 1991). In addition, half of molecular hydrogen 
mapped in CO (1.5 $\times 10^{10} M_{\odot}$ H$_{2}$) is located 
outside of the central 2.5 kpc (Sargent, Sanders, \& Phillips 1989).
And furthermore, NGC 3256 is the galaxy with the more 
extended polarized system known (Scarrott, Draper, \& Stockdale 1996). 
The extended nature at near-IR and CO confirm that NGC 3256 is not a very 
advanced merger.
Some advanced mergers such as Arp 220 and Mrk 231 are known as ULIRGs 
(Rieke et al. 1985; Heckman et al. 1987; Sanders et al. 
1988a; Lipari et al. 1994; Smith et al. 1998). 
Consequently, NGC 3256 seems to be evolving to an ULIRG, or even an elliptical
or radio galaxy.

Throughout the paper, a Hubble constant of H$_{0}$ = 75 km~s$^{-1}$ 
Mpc$^{-1}$ will be assumed. 
For NGC~3256 we adopted the distance of $\sim$37 Mpc (V$_{Syst}$ = 2817 
$\pm$15 km~s$^{-1}$; \S 3.2.1), and thus, the angular scale is 1$'' 
\approx$177 pc. (for an assumed Hubble constant of H$_{0}$ = 75 km~s$^{-1}$ 
Mpc$^{-1}$). 

\newpage

\section{OBSERVATIONS AND REDUCTIONS}

The optical and near-IR observations were obtained at Bosque Alegre, CASLEO,  
CTIO and ESO Observatories with the 1.54 m, 2.15 m, 1.0 m and 3.5 m 
telescopes, respectively. 
In addition, HST--(WFPC2 \& NICMOS) and ESO NTT--SUSI (Super Seeing Imager)
archive images of NGC~3256 were studied. 

The HST--WFPC2 observations include broad-band images, using the 
filters F450W ($\sim$B Cousin filter) and F814W ($\sim$I); with
a CCD scale  of 0.046$''$ pixel$^{-1}$, in the PC. 
HST--NICMOS observations include near-IR images, using the NIC2 and NIC1
CCD-camera, with the filters F160W(1.60 $\mu$m, wide),
F187N(1.87 $\mu$m, narrow), F190N(1.90 $\mu$m), F212N(2.12 $\mu$m),
F215N(2.15 $\mu$m), F222M(2.22 $\mu$m, medium), F237M(2.37 $\mu$m) (Table 1). 

ESO NTT data  include V broad-band images (obtained with EMMI, ESO Multi-Mode
Instrument), and [O {\sc iii}] $\lambda\lambda$4959+5007 and
H$\alpha$+[N {\sc ii}] $\lambda\lambda$6548+6584 narrow--band images (SUSI). 
The NTT--SUSI and EMMI data (Table 1) have seeing of 0.7$''$ 
and 1.4$''$  FWHM; and CCD scale of 0.13$''$ and 0.44$''$  pixel$^{-1}$, 
respectively. 

Moderate resolution long-slit spectrophotometric observations were 
obtained at the 1.0 m telescope of CTIO 
during two photometric nights in  March 1990 (Table 1).
The observations were made using a slit width of 1.5 arcsec, 
which gives an effective resolution of $\sim$8 \AA\ ($\sim$300 km s$^{-1}$)
and a dispersion of 
130 \AA\ mm$^{-1}$ covering the wavelength range $\lambda\lambda$3600--6900 \AA.
The seeing was approximately 1.5$''$  (FWHM).

Images and spectrophotometric observations of NGC~3256
were taken at CASLEO, San Juan (Argentina). 
Long-slit spectroscopic observations with moderate resolution were 
obtained with the University of Columbia spectrograph (see Lipari et al. 1997)
at the 2.15 m Ritchey-Chr\'etien telescope of CASLEO 
during two photometric nights in  March 1997 (see Table 1).
The observations were made using a slit width of 2.0 arcsec, 
which gives an effective resolution of $\sim$7 \AA\ ($\sim$280 km s$^{-1}$)
and a dispersion of 
130 \AA\ mm$^{-1}$ covering the wavelength range 
$\lambda\lambda$4100--7500 \AA.
Broad-band  CCD imaging observations were obtained in 
1997 March (see Table 1). Images through B, V and I filters were also obtained. 
A TEK 1K chip with a scale of 0.813$''$ pixel$^{-1}$ was 
used. The observations were obtained in photometric conditions and
the seeing was approximately 1.8--2.5$''$ (FWHM).

Extensive long-slit spectroscopic observations were 
obtained at the 1.54 m telescope of Bosque Alegre Station of Cordoba Observatory
using the Multifunctional Integral Field Spectrograph (Afanasiev, Dodonov \& 
Carranza 1994; Diaz et al. 1999) during 18 nights between April 1997 and 
March 1999 (see Table 1). 
The observations were made mainly using a slit width of 1.0 
arcsec and 1200 l/mm grating, 
which gives an effective resolution of $\sim$90 km~s$^{-1}$ and a dispersion of 
40 \AA\ mm$^{-1}$ covering the wavelength range 
$\lambda\lambda$6400--6900 \AA.
In order to have accurate spatial positions for the velocity
determinations in each extracted spectrum, zero order double imaging of object
plus slit before and after the spectroscopic exposures were used.
Spectrophotometric observations were also obtained using a 300 l/mm grating, 
whit an effective resolution of $\sim$330 km~s$^{-1}$, and 
covering the wavelength range $\lambda\lambda$4800--6900 \AA. 
The seeing was approximately 1.4--2.5$''$  (FWHM).
A Thomson 1K chip with a scale of 0.38$''$ pixel$^{-1}$ was used. 
The direct image mode gives a field of 5.7$' \times$5.7$'$.

UV--IUE/SWP\altaffilmark{7} and near IR archive spectra  (covering 
$\lambda\lambda$ 1200--2000 and $\lambda\lambda$ 7000--10000 \AA, 
with resolutions of $\sim$8 and 5 \AA, 
respectively) were obtained from the ``Starburst Galaxies Spectra Catalog" of 
Storchi-Bergmann, Kinney \& Challis (1995).

\altaffiltext{7}{IUE is the International Ultraviolet Explorer satellite;
and SWP is Short-Wavelength Prime (spectra).} 

The IRAF\altaffilmark{8}, software packages were used to reduce the 
imaging and spectrophotometric data (obtained at CTIO and CASLEO). 
Bias and dark subtraction and 
flat fielding were performed in the usual way. Wavelength calibration of the 
spectra was done by fitting two dimensional polynomials to the position 
of lines in the arc frame. The spectra were corrected for atmospheric 
extinction, galactic reddening, and redshift. The spectra and images 
were flux calibrated using observations of standard stars from the samples 
of Stone \& Baldwin (1983) and Landolt (1992). 

\altaffiltext{8}{IRAF is the imaging analysis software facility
developed by NOAO}

The kinematical observations obtained at Bosque Alegre (i.e., more than 25 
position angles of long-slit frames --mainly centered in the main optical 
nucleus-- from which were extracted approximately 1000 
individual spectra of 0.76$''$) were reduced and analyzed using mainly
SAO\altaffilmark{9}  and ADHOC\altaffilmark{10} software packages 
(see Diaz et al. 1999, for references and the velocity field reduction). 
The emission lines were measured and decomposed using Gaussians profiles 
by means of a nonlinear least-squares algorithm described in Bevington (1969). 

\altaffiltext{9}{SAO is the imaging analysis software developed by the 
Special Astrophysical Observatory, of USSR Academy of Science}

\altaffiltext{10}{ADHOC is the imaging analysis software developed by 
Marseille Observatory} 

\newpage

\section{RESULTS}

\subsection{Optical and Near--IR Structures}
\vspace{5mm}

CASLEO $B$ broad--band image, for all the extension of the 
merger, is displayed in Fig. 1a.
ESO NTT--EMMI $V$ broad--band of the central and external regions
of NGC 3256 is shown in Fig. 1b.
ESO NTT--SUSI high resolution narrow--band image in H$\alpha$+ 
[N {\sc ii}]$\lambda\lambda$6548+6584 of the central area 
is presented in Fig.\ 1c. 
The values of the seeing were 1.8$''$, 1.4$''$ and 0.7$''$, respectively.

HST--(WFPC2 \& NICMOS) broad/mid--band and color images
are presented in Figs.\ 1d, 1e and 1f (for the filters F814W,
F450W/F814W and F237M-NIC2).

This galaxy consist of a main body (r $\sim$9 kpc $\sim$60$''$), 
two extended tidal tails with a total extension of $\sim$80 kpc, and  
two faint external loops. 
In the central (r $\sim$5 kpc $\sim$30$''$) and nuclear 
(r $\sim$2 kpc $\sim$10$''$) 
regions several bright optical and IR knots are the prominent structures.
The general properties of NGC 3256 and luminous IR ongoing mergers are
presented in Table 2.

\vspace{5mm}

\centerline{{\it 3.1.1 The Nuclear and Central Regions (r $\leq$ 30$'' 
\sim$5 kpc)}}

\vspace{5mm}

The central regions of NGC 3256 show  a distorted 
ellipsoid shape with an extension to the south west 
(see Figs. 1b, 1c and 1d).
Inside of this ellipsoid there is a bright knotty structure 
that shows a very unusual triple  spiral arm morphology. 
The ESO NTT--SUSI and HST--WFPC2 images (with the filters
H$\alpha$+[N {\sc ii}] $\lambda\lambda$6548+6584 and F814W;
Figs. 1c and 1d) show that these arms are clearly three asymmetrical 
spiral arms; which projected in the sky -at low resolution-
look like rings or ``chaotic" knotty structures (Sersic 1959).

We call these three asymmetrical spiral arms as I, II and III (see Fig. 1c);
which start in the three main nuclear knots (Figs. 1d and 1f).
The arm I (which emanate from the main optical nucleus) shows the more
extended morphology, and a secondary arm in the north area.
The arm II, probably the nearest arm, shows  strong blue colors (see Table 3)
and is probably connected to the optical blue-knot located at 5.7$''$ to the 
east, from the main optical nucleus.
The arm III is partially obscured by dust and  is probably connected to the 
mid-IR knot located at 5$''$ to the south, from the main optical nucleus.
In the areas of intersection of the arms, Fig. 1c shows clearly that: 
(i) the arm II is behind the arm I; and (ii) the arm 
III is behind the arm II. Therefore, these structures/arms are in three
different planes.

In relation to the main nuclear knots (see Figs. 1c, 1d and 1h),
two prominent compact optical knots plus
a third compact mid-IR source have been previously studied. And, they are:
(i) the stronger optical knot or the main optical nucleus, that we call 
region 1; 
(ii) a similar compact knot located 5.7$''$ ($\sim$1 kpc) to the 
east from the optical nucleus (region 2); and 
(iii) an obscured optical knot detected only at wavelengths $\lambda \geq $ 3.75 
$\mu$m (region 3). 
These three knots are probably the nuclei of the original galaxies that 
collided (see Zenner \& Lenzen 1993; Moorwood \& Oliva 1994; Norris \& 
Fobes 1995; Kotilainen et al. 1995). 

In particular, 
regions 1 and 2 show giant H{\sc ii} regions properties with blue colors and 
starburst spectra (see  \S 3.3). 
The [O {\sc iii}]$\lambda\lambda$4959--5007 and $B$ band images show that 
region 2 has strong [O {\sc iii}] and blue emission, and these properties are 
also confirmed by the spectroscopic study (see \S 3.3 and Fig. 3b).

Using the NTT and HST data we 
study and compare the structure of the main optical nuclear knots and 
giant H{\sc ii} complexes. 
In particular, we are specially interested in the structure of the two main
optical knots/nuclei (regions 1 and 2).
The NTT H$\alpha$ and [O {\sc iii}] images show that the two knots have similar
morphology: a compact object ($\sim$ 0.9$''$ FWHM) surrounded by an extended 
envelope (r $\sim$ 3$''$). 
The HST--WFPC2 images show  new and more detailed structures, 
specially inside of the two optical NTT compact knots. In particular, for
the nuclear region 1, Figs. 1d and 1g show: 
(i) the nuclear envelope has a spiral--disk morphology of 
r $\sim$ 3$''$ $\sim$ 500 pc of radius, at face-on position; and 
(ii) even the NTT'compact main nuclear object is an extended structure
with also spiral--disk morphology, inside a radius of $\sim$ 0.9$''$
(Fig. 1g), plus two small central knots.
The two  spirals--disks are a continuation of the more external spiral 
arm I (this is evident in the continuum  and color images of the HST data:
Figs. 1d,e); however there are
several obscured areas  -probably- generated by the presence of dust.
The HST observations of region 2 (Figs. 1d and 1h) show:
an envelope with clumpy spiral structure (r $\sim$3$''$; which is less
clear than in the region 1, probably due to the high inclination of the
spiral--disk), and an extended structure plus two small central knots
{inside a radius of $\sim$ 0.9$''$; Fig. 1g).

Figs. 1g and 1h display at scale of sub arc--second,
the structure of the two main optical knots/nuclei (region 1 and 2).
These contours show relatively similar morphology for both regions: 
(i) a main central knot, which is clearly elongate (63 pc $\times$ 30 pc) for the case 
of the region 1, and it is compact for the region 2 (with a r $\sim$40 pc); 
(ii) a small compact knot (which is very close to the main central
knot, in each regions; with a r $\sim$16 pc);
(iii) and extended spiral--disk emission, probably the beginning of the asymmetrical 
spiral arms (which is also more resolved in the region 1).
These nuclear structures show all blue colors in the crude color-map of Fig. 
1e., and also in our BVI photomety (Table 3, obtained from CASLEO calibrated
images, with moderate spatial resolution).
In the \S 4.3 we discuss the properties of the
main blue elongate structure, found in the main optical nucleus of NGC 3256,
as a probable blue luminous star cluster (similar to those found by Shaya et
al. 1994, in the main/west nucleus and in the circumnuclear regions, of Arp 220).
Another interesting point to consider is the properties of the
small compact knots detected in each optical nucleus: the measured
sizes are the same to those proposed for young globular star
clusters (see Lutz 1991), but also for the accretion radius of  a low mass
nuclear black hole (of 10$^7$ M$_{\odot}$; Heller \& Shlosman 1994).

The optical obscured region 3 shows for wavelength $\lambda \geq $ 3.75 
$\mu$m ($\sim$L$'$ filter) the typical properties of a nuclear region; 
i.e., compact and strong continuum emission, typical multiwavelength color 
index, etc. (see Kotilainen et al. 1995;  Zenner \& Lenzen 1993; Moorwood \& 
Oliva 1994; Norris \& Fobes 1995). 
The HST--NICMOS images (which reach only $\lambda \leq$2.5 $\mu$m) show that 
region 3 has weak near-IR emission, and it appears to be  
connected to the arm III (Fig. 1f). 
In general, these short exposure HST--NICMOS images show the strongest 
circumnuclear knots; and Fig. 1f display mainly the regions 1 and 3, and
the three arms (in a field of 19$'' \times$19$''$; for NIC2).

Table 3 includes the BVI photometric properties of the main knots and structures. 
We comment here some general properties of these nuclear 
and circumnuclear regions: 
(a) regions 4, 6, 8, and  9 are the strong blue knots
in the arm II (and close to region 2);
(b) region 5 is a strong H$\alpha$ knot at the end of arm I
(called as region ``X" by Feast \& Robertson 1978); 
(c) regions 7a and 7b are the two main H$\alpha$ knots in the arm III
(called as region ``S" by Feast \& Robertson 1978 
and ``A" by Laubert et al. 1978).
The region 10 (at 7.3$''$ north and 5.0$''$ west, from the main optical
nucleus) is a strong H$\alpha$ knot in the arm I; and
it is close to the 3rd. illumination source of polarimetry 
(at 6.1$''$ north and 3.5$''$ west), 
detected by Scarrott et al. (1996).
The spectra and physical conditions of some of these regions are 
presented and analyzed in \S\S 3.3 and 4.4.

Other, interesting  central regions -see Table 3- are: 
(a) regions 11, 13, 14, and 15
are bright knots inside of the arm I (in the north area);
(b) region 12 is a very blue knot inside of a secondary arm, which is part of 
the main arm I;
(c) region 16 is a strong H$\alpha$ knot, located in a secondary or probable
obscured fourth arm (see for detail the next \S);
(d) region 17 is a knot in the disrupted SE external arm;
(e) region 18 is a H$\alpha$ ``isolated" knot,
located to the north of region 7b.

From Table 3 it is interesting to note that there is a clear continuity 
in the photometric properties of each  spiral arm; 
and also with each corresponding nucleus.
In particular, in the arm I there is continuity in colors 
 between the regions 1, 10, 11, 12, 13, 14, 15 and 4 (see Table 3). 
In the arm II, the continuity is detected between regions 2, 4, 8, 9 and 6;
and in the arm III the continuity appears
between the regions 3, 7a, and 7b. On the other hand,
the strong blue colors found in the arm II and region 2, could be related 
to intrinsic properties plus the fact that this is probably the nearest
arm (see Figs. 1c and 1d)

The [O {\sc iii}]$\lambda\lambda$4959--5007 narrow-band image shows -in general-
similar morphology to the H$\alpha$ image (Fig. 1c): i.e.,
extended triple asymmetrical spiral arms structure which probably 
emanate from three different nuclei. However, there are some differences, in 
particular the arm II and the region 2 show strong [O {\sc iii}] emission.

The 3 and 6 cm radio observations of the central region of this 
merger (Norris \& Forbes 1995) show clearly extended continuum emission with
two strong compact peaks at the position of regions 1 and 3; 
plus three weaker/secondary peaks at the position of regions 2, 7 and 10.
They associated the two main compact radio continuum peaks to non-thermal
emission from remnant of supernovae (SN), which would occur 1 every 3 years
(and they suggested that the two main compact knots are the original nuclei 
of the two galaxies that collide). 

Finally, we note that the continuum 
(i.e., B, V, I, J, H, K$'$, L$'$, 3-6 cm) and the emission lines 
(i.e., [O {\sc iii}]$\lambda\lambda$4959+5007, H$\alpha$+[N {\sc ii}]
$\lambda\lambda$6548+6584, Br$\gamma$, 
[Fe {\sc ii}], CO(2-1)) 
at different wavelength show {\it extended and similar morphology},
in a radius of $\sim$5 kpc 
(see our Figs. 1b, 1c, 1d and 1f; and also Norris \& Fobes 1995; Kotilainen
et al. 1995; Aalto et al. 1991; Moorwood \& Oliva 1994).
This fact is in good agreement with the suggestion that NGC 3256 
is a relatively advanced merger, which is still far from the state of: 
(i) fusion of the nuclei, of the original galaxies that collide, and 
(ii) high concentration of ionized and molecular gas in the nuclear region.
Therefore, interesting process could be still observed in this merger,
for example inflow of the ionized H$\alpha$ gas. 
Furthermore, this fact is clearly indicative that the same process: i.e.,
starburst, is the 
dominant source of energy at optical, IR, millimeter and radio wavelength.

\vspace{5mm}
\centerline{{\it 3.1.2 The Outer Regions (r $\geq$ 30$'' \sim$5 kpc)}}
\vspace{5mm}

In the  external regions (Figs. 1a and 1b) we study first
the extended envelope (r $\sim$9 kpc), which shows a rhomboidal shape and 
include -in the southern part- two probable disrupted arms.
We found, for r $\geq$ 30$''$ {\it ``radial 
filaments"} and dust lines, which emerges from the central region
(see Fig. 1b). In 
this area Scarrott et al. (1996) measured very extended polarization
(r $\sim$7-9 kpc) attributed to a dust reflection and starburst/galactic--wind. 
These results are consistent with the detection of outflow in the nuclear 
starburst and extended emission regions (see the  \S 3.2.2).

On the other hand, in the southern part of the external  
regions there are only few strong emitting structures (see Fig. 1b).
This fact could be explained
by the presence of strong reddening in these areas (Scarrott et al. 1996). 

In the obscured southern area --of the 
central and external regions-- we detected in H$\alpha$, [O {\sc iii}]
and continuum images (Figs. 1c and 1d) a secondary or a probable
fourth asymmetrical arm, strongly obscured by dust. Which also 
emanate from the nuclear area, probably from one of the strong H$\alpha$ or 
obscured nuclear knots (e.g., region 2, 3, or 4).
And, this secondary or arm-IV reach the compact H$\alpha$ knot located
at 6$''$ south and 6$''$ east (region 16, in Table 3), 
and probably also an extended H$\alpha$ knot located at 18$''$ south and 
16$''$ west (region 19, in Table 3). 

In the outer part of the main body of this merger there are two
disrupted externals arms; which are probably connected to external loops
(Figs. 1b and 1a).
In particular, the SW external arm shows a more extended
morphology and a clear connection to the SE loop (see Fig. 1b). And,
the region 17 (which is inside of the SE disrupted external arm) shows
clear photometric continuity with the SW loop (Table 3).

Table 3 include also the BVI photometric properties of several interesting 
external regions. We comment here general properties of these areas: 
(a) region E-tail/1, E-tail/2, E-tail/3, and E-tail/4
are the four main/strong blue knots inside of the E tidal tail; and
(b) region W-tail/1, W-tail/2, W-tail/3, and W-tail/4
are the four main blue knots inside of the W tail.

In both tails several condensations show blue colors; and 
it is interesting to note that the more intense and bluer regions are located 
at the West tail (see Table 3). 
These blue regions are probably associated to star-formation process, as has 
been observed in similar major mergers IRAS 19254--7245, NGC
4038/39, and Mrk 231 (by Mirabel, Lutz \& Maza 1991; 
Colina et al. 1991; Mirabel, Dottori \& Lutz 1993; Lipari et al. 1994).

In conclusion,
the {\it morphological  results}, combined with the spectroscopic 
observations (presented in \S 3.3) are clearly compatible 
with an extended massive starburst.
And, these results are in good agreement with previous studies, obtained 
at different wavelength (specially with IR observations; see 
Graham et al. 1984; Joseph 1991).

\subsection{The Kinematics (of the Ionized Gas)}
\vspace{5mm}

NGC 3256 is probably the nearest ``advanced" major--merger known,
however the main kinematical properties of this system remain undefined. 
Furthermore, different works (e.g., Feast \& Robertson 1978; Sargent et al. 
1989; Scarrott et al. 1996) suggested that a detailed kinematics study of 
this merger is required. 

Our kinematical observations consist of more than 1000 individual 
spectra, which were extracted from long-slit frames obtained at $\sim$25 
position angles (with an instrumental width of $\sim$90 km~s$^{-1}$ FWHM,
in the sky emission lines). 
We measured the velocities from the centroids of the stronger emission 
lines H$\alpha$, [N {\sc ii}]$\lambda$6584 and [S {\sc ii}]$\lambda
\lambda$6717--6731, fitting gaussians. 
The long-slit positions were obtained mainly passing through the main optical
nucleus (each 10$^{\circ}$--20$^{\circ}$), in order to use 
the nucleus velocity as reference, i.e., as the zero value in the velocity
 field (VF). Therefore, 
each point in the VF is the result of the determination: 
(a) the velocity obtained from high quality spectra (S/N$\sim$10--30); and 
(b) the ``relative" position, which is accurate on the scale of both VFs 
(0.5$''$ and 1$''$) presented here.

Using high resolution data, for kinematical studies, an important point  
is the analysis of multiple components, in each  emission lines (H$\alpha$,
[N {\sc ii}]$\lambda$6584, [S {\sc ii}]$\lambda$6717--6731). 
For Cen A, NGC 1052, 4550 and 7332 this type of study --of multiple 
components-- was made by Bland, Taylor, \& Atherton (1987); Rubin, Graham, \& 
Kenney (1992); Plana \& Boulesteix (1996). 
In NGC 3256, first we study only the main emission line component; 
which is very clear/strong in the central region. However,
in the outer regions the presence of multiple components -with
similar weak strength- required a detailed study (see \S 3.2.3).

In addition, for this set of data -which cover practically all the main 
structures and reach the more external part of the system- we also 
measured the fluxes and FWHM of the emission lines; in order to study 
the presence of different kinematical components, and to complement the study
of the physical condition (see \S 3.3).

The inclination value adopted as a first approximation in our kinematics
study was i = 40$^{\circ}$ (from Feast \& Robertson 1978); and the final
adopted values was i = 42$^{\circ} \pm$5$^{\circ}$, which minimized the sum 
of the squares of the residual velocity values (in the regions where the
circular motion prevails).  And, we assume that the 
nearest side of the merger is the north part, i.e., the less obscured area.

\vspace{5mm}

\centerline{{\it 3.2.1. The Kinematics in the Nuclear and Central Regions
(r $\leq$ 30$'' \sim$5 kpc)}}

\vspace{5mm}

Figs. 2a and 2b show  the H$\alpha$ velocity field -color-map and contours-  
for the central 40$''\times$40$''$ ($\sim$7 kpc$\times7\,kpc$, 
and r$_{max}$ $\sim$30$'' \sim$5 kpc), with a spatial resolution/sampling
of 1$''\times$1$''$. The errors vary from approximately
$\pm15\,$km~s$^{-1}$ in the nuclear and central regions (where the emission
lines are strong), to $\sim$30 km~s$^{-1}$ for the weakest lines away
from the nuclear areas.
A $\sim$35$\%$ of the grid points of the velocity field 
were obtained by linear interpolation of observed radial velocity,
mainly in the outer radii (the central regions are oversampled).

We observed in the nuclear region of the VF (r $\leq$ 5$'' 
\sim$1 kpc) the typical shape of circular motion: i.e.,
the {\it ``spider" shape} (Fig. 2b). And, we determine the precise location of 
the kinematical center of rotation 
using different methods (i.e., interception of isovelocity lines, square 
root of nuclear rotation curve residuals, etc.).
This center is located  3$''$ S and
1-2$''$ W from the main optical nucleus; i.e., very close to the line 
that connects the main optical nucleus and the obscured nuclear knot
(region 3).

Furthermore, we have also studied the velocity pattern in the neighborhood
of the kinematical center and the main optical nucleus (r $\leq$ 4$'' 
\sim$1 kpc) using the plot VR vs cos(PA-PA$_0$), where PA is the position
angle of the radial velocity point and PA$_0$ is the position angle of the 
kinematical line of nodes/major-axis.
We found rotational pattern around both areas.

The extended H$\alpha$ color and isovelocity maps (Figs. 2a,b) 
has an amplitude of $\sim$250 km s$^{-1}$, and show 
interesting and complex substructures: 
(i) in the negative velocity region (blue shifted) there are two defined subregions,
one close to the arm III (SW region, with a strong blue peak of $\Delta$ 
V $\sim$--150 km s$^{-1}$) and other close to the 3rd. source of polarization 
(Scarrott et al. 1996; in the NW area); 
(ii) in the positive velocity region there are three clear 
subregions (but they can not be easily associate to any particular structure).
The main region of high negative values shows  continuity to the external 
region and specially to the west tail. 
There is also continuity of weak positive values, in the north part of the VF,
in the direction to the east--north tail.

The systemic velocity (V$_{Syst}$) was determined using different methods. 
First, for each slit position passing through the main optical 
nucleus the V$_{Syst}$ was
determined as the mean value of the external positions of the corresponding
velocity curve. In this way the resulting mean V$_{Syst}$ --for all the PA and 
corrected by heliocentric motion-- was  2815 $\pm15\,$km~s$^{-1}$.
In addition,
we measure the velocity of the kinematical center as the velocity of the spider 
center (using the mean velocity of the main optical nucleus, 
V$_{Region 1} =$ 2823 $\pm12\,$km~s$^{-1}$), and we found V$_{Syst} =$ 2819 
$\pm12\,$km~s$^{-1}$. Therefore, the mean value for the velocity of the system 
is V$_{Syst} =$ 2817 $\pm15\,$km~s$^{-1}$. 

On the other hand,
the velocity obtained for the CO kinematical center was 2800 
$\pm30\,$km~s$^{-1}$ (see Aalto et al. 1991; and \S 4.3).
Previously, Feast \& Robertson (1978) found for the system and region 7a
velocities of 2823 $\pm25\,$km~s$^{-1}$ and 2710 $\pm25\,$km~s$^{-1}$,
respectively.
Zenner \& Lenzen (1993) obtained for the main optical nucleus a value of 2818 
$\pm18\,$km~s$^{-1}$. Therefore,  the velocity values obtained in this work 
are in good agreement with those reported previously. 

The three non coplanar arms, described in \S 3.1, can be traced through the
residuals of the VF lobes, and one of them (arm II) has clearly an opposite 
sense of rotation; in agreement with the general H$\alpha$ and broad-band 
morphology (Figs. 1c and 1d). 

Three superposed rotation curves (RCs), we found from the extraction in the
VF of a wide aperture cone ($\theta$ $\sim$80$^{\circ}$) at
PA$\sim$90$^{\circ}$.
Then, we separated these RCs using extraction of small aperture 
cones $\theta$ of 30$^{\circ}$, around the PA
$\sim$55$^{\circ}$, $\sim$90$^{\circ}$, and $\sim$130$^{\circ}$ (Fig. 2c). 
These RCs could be associated to the three main lobes observed in the 
positive and negative regions of the VF; and show the typical
``sinusoidal shape" of merger galaxies (Schweizer 1982; Rubin \& Ford 1983;
Zeph 1993). In \S 4.3 we discuss the properties of these RCs.

Moorwood \& Oliva (1994), from Br$\gamma$ velocity map, already suggested 
the presence of more than one kinematical axis, in NGC 3256. 
And, they proposed a major kinematical axis at PA
$\sim$65$^{\circ}$ (which is close to the PA $\sim$55$^{\circ}$, of one of
our RC).
Previously, Feast \& Robertson (1978) and Aalto et al. (1991) suggested a
major  kinematical axis at PA $\sim$110$^{\circ}$, and $\sim$90$^{\circ}$,
respectively. 

On the other hand, for the nuclear region (r $\leq$5-7$'' \sim$1 kpc; 
where we detected circular motions), we obtained the central
part of three RCs from a nuclear VF with 0.5$''$ sampling, at
PA$\sim$55$^{\circ}$, $\sim$90$^{\circ}$ and $\sim$130$^{\circ}$ (Fig. 2d).
The center for these RCs is located very close to the main nucleus: at 1$''$
south and 2-3$''$ west.
These three high spatial resolution RCs
show clear sinusoidal curves superposed to the general RCs, with a symmetric 
``fall/break" at $\sim$ $\pm$4$''$ (r $\sim$700 pc); and we obtained also a
similar result using the extended VF with 1$''$ resolution (but, less clearly).
In addition, we study the presence of this feature in all the
nuclear VF: we extracted
RCs using a step of 20$^{\circ}$; and
we found similar superposed sinusoidal curves shape in the PA range of
40$^{\circ}$--130$^{\circ}$. And, this shape in the RCs is very similar to
that found in the nuclear velocity curve of the Seyfert 1 
galaxy NGC 6860. In this AGN, the superposed curve was associated to the possible 
presence of inflow (Lipari, Tsvetanov \& Macchetto 1993b). 

In addition, the region with a probable inflow (r $\leq$ 4$''$) 
is approximately where the HST data (\S 3.1) show a spiral--disk structure 
around the main optical nucleus. 
A similar spiral structure was found by Colina et al. (1997) 
--from HST observations-- in the Liner nucleus of NGC 4303; and they suggested
that this structure is probably related to the fueling of the AGN (i.e., 
inflow process). Finally, also the VF shows in the nuclear and central regions 
several isovelocity contour suggesting the presence of inflow, i.e., several
contours display ``clear elongation and asymmetry" pointing directly to the
main optical nucleus (see Figs. 2a,b). 

We measured also the velocities of the [N {\sc ii}]$\lambda$6584  and
[S {\sc ii}]$\lambda$6717--6731 emission lines, for each PA. And we detected
only small differences of $\sim$15--20 km~s$^{-1}$, in
relation to H$\alpha$ data (in the nuclear and circumnuclear regions). 

\vspace{5mm}
\centerline{{\it 3.2.2. The Kinematics of the Outflow and Multiple-Components}}

We found direct evidence of outflow in the main optical nucleus (region 1)
of NGC 3256: practically all the spectra show -in the nuclear region-
blue components in the H$\alpha$, [N {\sc ii}]$\lambda$6584 and
[S {\sc ii}]$\lambda$6717 emission lines (see Fig. 2e). And,
we detected interesting variations of this blue component for different 
PA. In order to check these variations we observed the same PA using 
different instrumental configuration, and we obtained the same result: 
there is a blue/outflow component for practically all the PA, however in the 
interval of PA 45$^{\circ}$--80$^{\circ}$ this component practically
disappear or is very weak.

Specifically,
for PA without outflow component (e.g., PA 45$^{\circ}$) we made the test of 
increase the long-slit aperture (from 1$''$ to 3$''$), and in all the case
-for this PA- the blue component was not detected. A second test, was performed
for PA with  outflow component (e.g., PA 110$^{\circ}$) moving the long-slit
position very close to the nucleus, we detected the blue component
only when the long-slit was exactly at the center of the nucleus. Therefore,
we conclude that the outflow is strong mainly in the nuclear area, and 
without --or very weak-- component at PA 45$^{\circ}$--80$^{\circ}$ (i.e., 
bipolar outflow, with an opening angle $\theta$  $\sim$140$^{\circ}$).

We note that this blue outflow component was clearly detected only using 
data of relatively high resolution ($\sim$90 km~s$^{-1}$ FWHM). And, 
the velocity of this nuclear outflow is V$_{Nucl.OutFlow} \sim$
350 km~s$^{-1}$, and the FWHM is $\sim$130 km~s$^{-1}$.
In \S 4.5 this component  will be associated  to the nuclear starburst.

On the other hand,
in order to study the outflow component in the central and external  
regions, long exposure spectroscopic observations were performed  
(i.e., add-spectra of $\sim$3-4 hs.; see Table 1).  
In particular, at two selected position angles: 
PA 18$^{\circ}$  and 70$^{\circ}$ (close to the minor kinematic axis and to
the region 7, respectively).
Fig. 2f shows the radial velocity vs. axial distance (from the nucleus) 
for ``the main and the outflow components", at PA 18$^{\circ}$ and
for H$\alpha$, N[ {\sc ii}] and [S {\sc ii}] emission lines.  
We detected extended-outflow --in the
emission lines [S {\sc ii}]$\lambda$6717, [N {\sc ii}]$\lambda$6584  and 
H$\alpha$-- from the main optical nucleus to r  $\sim$30-40$''$. However,
we detected and measured the outflow component in all the range
(r $\leq$30$''$ $\sim$5 kpc; Fig. 2f),
only in the line [S {\sc ii}]$\lambda$6717.
Which could be explained by a combination of two facts: (a) strong
flux in the [S{\sc ii}] outflow component, and (b) the relatively easy
detection and measurement of the [S{\sc ii}]$\lambda$6717 outflow component,
since it is located in the blue-border
of the blend [S {\sc ii}]$\lambda$6717+6731 (main component).
This extended outflow, reaching a r $\sim$5--6 kpc, was already suggested 
by polarimetry study (Scarrott et al. 1996). 
   
We observed for the galactic--wind in NGC 4945 (the nearest starburst and 
Seyfert nuclei, Lipari et al. 1997) similar relation in the strength of
these three emission lines: strong [N {\sc ii}] and [S {\sc ii}], and weak 
H$\alpha$ (see Lipari et al. 1997: their Figs. 3b and 3c). 
For NGC 3256, we measure in the extended outflow component, the following
values of the emission line ratios:
(i) [S {\sc ii}]${\lambda 6717+31}$/H$\alpha$ in a range 0.7--1.3; and 
(ii) [N {\sc ii}]$\lambda$6584/H$\alpha$ in a range
0.8--1.6. These values are clearly consistent with 
shocks driven into clouds accelerated outward by a starburst with galactic--wind
(see \S 4.5; HAM90: their Fig. 14; Lipari et al. 1994, 1997). 

Finally, another interesting result is the presence of multiple components
in this system. 
In the central regions we detected multiple components only in few
areas, specifically close to the arm III (the area of high negative velocity
values and the region 7), in the south-west central
area (r $\geq$ 10$''$). In addition, in these regions the FWHM of the main
component --of the emission lines-- shows a clear
increase in their values, reaching a FWHM of $\sim$260 km s$^{-1}$ 
(see Figs. 4g and 5d); and we observed clear double peak.
In the more external areas (r $\geq$ 5 kpc), we found also
clearly double and multiple weak components, specially in the north-east 
regions.

\vspace{5mm}
\centerline{{\it 3.2.3. The Kinematics in the Outer Regions (r $\geq$ 30$'' 
\sim$5 kpc)}}

\vspace{5mm}

Fig. 1c shows clearly 
that only in the nuclear and central area (r $\leq$ 5 kpc) there is strong 
H$\alpha$+[N {\sc ii}] emission. In the external areas,
the flux and S/N of the emission lines decrease (S/N $\sim$ 2--5), and 
the errors increase to $\sim$30--50 km~s$^{-1}$.

We detected --at this low S/N-- some interesting results in these outer areas:
(i) the region of high negative velocity values shows continuity (in their
values), in the direction to the W tail; 
(ii) observations of large exposure time, along the tails, suggest that 
in NGC 3256 the H$\alpha$ ionized gas present --in general-- a similar 
velocity behavior of the atomic HI gas in the merger NGC 7252 
(by Hibbard et al. 1994): i.e., the H$\alpha$ ionized-gas are swinging through 
space in the same sense as the rotation of the central ionized-gas. 

Specifically, in the W tail the velocity increase from -100 km s$^{-1}$ at
the base of the tail (which is located at $\sim$22$''$ west and 16$''$ south,
from the region 1) to -130 km s$^{-1}$ at 13-18$''$ from the base of the tail;
and then the velocity gradually decrease.
In the E tail the kinematical behavior is similar, however the positive values
are smaller in amplitude, to those obtained in the W-tail.

Finally, is interesting to note that
in the external regions (r $\geq$ 20$''$) there are some  kinematical
differences ($\sim$50--80 km~s$^{-1}$) mainly between [N {\sc ii}] and 
H$\alpha$ emission lines. 
In particular, Fig. 2f shows in the outer and also in the central regions 
(close to the arm III and regions 7a, and 7b),
clear velocities diferences between the main components, in these emission line.

\subsection{Optical Spectrophotometry}
\vspace{5mm}

There are several IR spectophotometric studies of NGC~3256, however 
there are only {\it few general studies at optical wavelengths} (see 
Laubert et al. 1978; HAM90; Aguero \& Lipari 1991). 
Sargent et al. (1989) already suggested that a detailed spectroscopic study
of the {\it ``optical knotty"} structure of NGC 3256 is required in order 
to determine clearly the nature of these knots/clumps. 

We have obtained at CTIO, CASLEO and Bosque Alegre optical long-slit 
spectroscopy (with a final instrumental resolution of $\sim$290 km~s$^{-1}$)
of the main structures at the nuclear and central region of NGC~3256, in
order to study the continuum and emission lines in the range
$\lambda\lambda$3600-7000.
In addition, we use the extensive long-slit spectroscopy obtained at Bosque 
Alegre mainly for kinematical propose (with instrumental resolution of 
$\sim$90 km~s$^{-1}$), in order to study the spatial variation of the emission 
lines in the spectral range $\lambda\lambda$6400-6900 \AA. 
These data covered practically all the extension 
of this merger galaxy (including the extended tails).

Finally, we also studied UV-IUE/SWP and near-IR archive data of NGC 3256;
obtained with moderate spectral resolution ($\sim$300 km~s$^{-1}$) and 
large aperture (10$''\times$20$''$; which include the main parts of the 
central region).

\vspace{5mm}

\centerline{{\it  3.3.1. Optical, UV and near-IR Spectrophotometry of Moderate 
Resolution}}

\vspace{5mm}

In this section we present results of optical, UV and near-IR 
spectrophotometry (of moderate spectral resolution: $\sim$290 km~s$^{-1}$
FWHM), for the main structures studied in \S 3.1. Which combined with the data 
obtained from photoionization codes (e.g., Ferland 1991) and
models of high metallicity giant H{\sc II} regions, massive star formation 
and shocks
(Garcia Vargas el al. 1997; Pastoriza et al. 1993; Garcia Vargas \& Diaz 1994; 
Garcia Vargas, Bressan, \& Diaz 1995a, 1995b) will allow us to determine 
the physical condition in the emitting gas and also in the main ionization 
sources. 
These results will be also compared with those obtained previously for 
the near and mid-IR wavelength.

First, we present in Tables 4, 5 and Figs. 3a-i, the results obtained from 
moderate-resolution  {\it optical} long-slit spectra  
for the main central regions 1 , 2, 4, 5, 6, and 7; plus UV and near-IR 
spectra of the circumnuclear area. 
From these data the main conclusions are the following: 
(i) all the observed spectra in the central area show similar 
characteristics, i.e., high metallicity giant H{\sc II} regions or
starburst (extended in a r $\sim$ 30$'' \sim$5 kpc; the same area of 
strong optical, IR, millimeter and radio continuum and emission lines);
(ii) the two compact nuclear regions --1 and 2-- show luminosities in H$\alpha$ 
compatible with nuclear starburst (see Table 4 and Kennicutt, Keel \& 
Blaha 1989);
(iii) in the central area only the regions 7a,b (in the arm III) show some 
differences in the spectra, i.e., a strong reddening in the continuum 
and the emission lines.

In particular,
the observed spectra in the three asymmetrical spiral arms and in the two main optical 
knots/nuclei (Figs. 3a-i and Table 4) show --practically all-- 
the typical features of {\it low ionization and 
relatively high metallicity giant H{\sc ii} regions} (Diaz et al. 1987, 1990, 
1991; Terlevich 1993; Garcia Vargas \& Diaz 1994) showing: very low 
[O {\sc iii}] ${\lambda 5007}$/H$\beta$ ratio ($\sim$0.5), and absent or 
very weak [O {\sc iii}]${\lambda 4363}$ and [N {\sc ii}]${\lambda 5755}$.

On the other hand, in order to study the stellar populations it is 
also important to analyze the {\it absorption spectra} of NGC 3256. 
In particular, the more clear ``signature" of massive
stars and starburst is the presence of absorption lines at the UV region 
of the spectrum (see Weedman et al. 1981; Leitherer 1991).
From the UV--IUE library of Seyfert and starburst galaxy (McQuade, Calzetti, 
\&  Kinney et al. 1995; Storchi-Bermann, Kinney \& Challis 1995) we have
obtained the UV spectra of NGC 3256, in the range $\lambda\lambda$ 1200-2000
\AA. This spectrum shows mainly, for the circumnuclear region
(10$''\times$20$''$; Fig. 3g), the typical starburst broad absorption lines:
Si{\sc iv}$\lambda$1400 and C{\sc iv}$\lambda$1550.
These  lines show profiles very 
similar to the prototype of starburst galaxy NGC 7714 (Weedman et al. 1981).
We measured the ratio of the equivalent width (EqW) of these lines 
and we found a value of EqW(Si{\sc iv})/EqW(C{\sc iv}) = 0.76$\pm$0.06,
using the spectral windows of
$\lambda\lambda$1380--1415 and $\lambda\lambda$1423--76, respectively
(these are the windows already used by Storchi--Bermann et al. 1995 and 
Robert, Leitherer \& Heckman 1993, for their analysis of UV-IUE data).
The EqW(Si{\sc iv})/EqW(C{\sc iv}) ratio is an important parameter in order 
to study the IMF, and our measurement is in agreement with the average obtained 
for starburst galaxies: 0.69$\pm$0.13 (Sekiguchi \& Anderson 1987; 
Scalo 1990)  and 0.83$\pm$0.1 (Robert et al. 1993; see \S 4.4).

It is important to note, that the two main ``broad" absorption lines
show --in the IUE spectra-- a strong blueshift of
V$_{Si IV}$ =  --750 $\pm$60 km s$^{-1}$ and 
V$_{C IV}$ = --1070 $\pm$40 km s$^{-1}$
(see Table 6). In agreement with the prediction
of starburst models with strong galactic--wind from massive stars;
since the Si{\sc iv} and C{\sc iv} absorption lines can be
formed by stellar winds with P Cyg profile (Robert et al. 1993), and also
by the ISM, and by the photosphere of B stars (in the case of Si{\sc iv}).
Furthermore, in the UV region NGC 3256 shows a composite spectra, 
showing features typical of starburst (e.g., NGC 7714), strong Si{\sc iv}, 
C{\sc iv} and He{\sc ii} lines; and 
also of IR galaxies (e.g., NGC 7552) with strong Fe{\sc iii}+Al{\sc iii} blend 
at $\lambda$1600--1630 and many weak absorption lines from young stars, 
plus a strong red UV--continuum (see \S 3.4). And, no shift is expected for
the  Al{\sc iii} and Fe{\sc iii} lines, since they are originate in the ISM
and in the photosphere of the stars (Robert et al. 1993).
 
Table 6 shows the equivalent width (EqW) --corrected by galactic reddening-- 
for the main {\it UV, optical and near-IR absorption lines} (from spectra 
obtained at CTIO, CASLEO, and IUE; Figs. 3a-f).
We measured for the region 1+2 --in the near-IR-- the EqW of the 
absorption triplet of Ca{\sc ii}, (Ca {\sc ii}$\lambda\lambda$8498, 
8542, 8662) $\sim$ 1.1, 5.4, and 3.4 \AA, respectively.  
And, these lines were not detected as emission features
(see the discussion of these results in \S 4.4).

\vspace{5mm}

\centerline{{\it  3.3.2 Optical Spectrophotometry with High Spectral/Spatial 
Resolution}}

\vspace{5mm}

From the relative high spectral  and spatial resolution 
(90 km~s$^{-1}$ and $\sim$0.8$''$ sampling)
long-slit {\it optical} spectroscopy, we study the 
spatial variation of the: (a) emission lines, (b) the 
physical conditions, and (c) the ionizing sources, in NGC 3256.
In particular, it is important to know the 
physical conditions in the nuclei, the spiral arms, the outflow component
and the more external regions.

First, we present in Figs. 4a-h
the spatial variation of the emission line ratios 
[N {\sc ii}]$\lambda$6584/H$\alpha$,  
[S {\sc ii}]${\lambda 6731}$/[S {\sc ii}]${\lambda 6717}$, 
[S {\sc ii}]${\lambda 6717+6731}$/H$\alpha$,
 and the FWHM of the H$\alpha$ emission line (FWHM(H$\alpha$)); for seven selected PA
(70$^{\circ}$/56$^{\circ}$, 155$^{\circ}$/18$^{\circ}$,
45$^{\circ}$/155$^{\circ}$, and 65$^{\circ}$/40$^{\circ}$).
These Figs. show that:
(i) the ratio [N {\sc ii}]$\lambda$6584/H$\alpha$ decrease from 0.60 in the 
main optical nucleus and 0.80 in several outer regions to 0.30 in the
circumnuclear areas;
(ii) the ratio [S {\sc ii}]${\lambda 6731}$/S {\sc ii}]${\lambda 6717}$ 
decrease from 1.08 in the main nucleus and 0.95 in the spiral arms to 0.6--0.70 
in the external areas; 
(iii) the ratio [S {\sc ii}]${\lambda 6717+6731}$/H$\alpha$ increases from 
0.20--0.30 in the central regions to 0.40-0.60 in the outer regions
(for PA inside of the outflow bipolar cone; e.g., PA 155$^{\circ}$, Fig.
4f); and it is constant --0.20--0.30-- for PA outside of this outflow 
area (e.g., for PA of 45$^{\circ}$, Fig. 4e);
(iv) the FWHM(H$\alpha$) increase from  the instrumental profile value
of $\sim$90 km s$^{-1}$ in the outer regions to $\sim$270 km s$^{-1}$ in the 
lobe with high blue velocity values and the arm III; and for
the two main optical knots (regions 1 and 2), the FWHM(H$\alpha$) reach
$\sim$220 km s$^{-1}$.

In Figs. 5a-d we also display detailed maps of emission line ratios
[N {\sc ii}]$\lambda$6584/H$\alpha$,  
[S {\sc ii}]${\lambda 6731}$/[S {\sc ii}]${\lambda 6717}$, 
[S {\sc ii}]${\lambda 6717+6731}$/H$\alpha$,
 and FWHM(H$\alpha$);
for the central region of NGC 3256 ($\sim$30$\times$30$''$, with 1$''$
spatial resolution). These maps were made, using the similar technique
described in \S 3.2, for the case of the velocity field map.
It is important to note that these maps were made for the main
component of the emission lines.

Figs. 5 show very interesting results, and we remark:

\begin{enumerate}

\item
In the central 20$''\times$20$''$ (r $\leq$13$''$ $\sim$ 2 kpc)
all the emission line ratios maps show
-for the main component- the typical values of giant H{\sc ii} regions
(see Terlevich et al. 1991; Kennicutt et al. 1989; and their references).

This result is in agreement with the fact that: this is the area of
strongest emission lines (Figs. 1c and 1d), which come mainly from the
nuclei and the asymmetrical spiral arms (where we detected ongoing massive
star formation process, see \S 3.3.1 and Figs. 3).

\item
The [S {\sc ii}]${\lambda 6717+6731}$/H$\alpha$ map shows 
for r $\geq$10$''$ the typical values of ionization by shock heating
([S {\sc ii}]${\lambda 6717+6731}$/H$\alpha$ $\geq$0.32; HAM90).

Furthermore, there are two main extended regions, where this ratio reach very
high values (in the range 0.40--0.60), and they are aligned with the same PA
of the outflow component axis (PA $\sim$150$^{\circ}$).
The other extended region that shows high values of this
[S {\sc ii}]/H$\alpha$ ratio is located close to the 3rd. source of
polarization (Scarrott et al. 1996).

\item
The [S {\sc ii}]${\lambda 6731}$/[S {\sc ii}]${\lambda 6717}$ map shows 
two strong lobes at the position of the two main optical nucleus: region 1
and 2 (reaching values in the range  1.00--1.10), with a elongation in the
area of the region 3.

And, there are also secondary peaks (reaching values of 0.94--0.98) in the
other circumnuclear giant H {\sc ii} regions: 4, 6, and 9. 

\item
The [N {\sc ii}]$\lambda$6584/H$\alpha$ map shows a complex structure, 
with high values in the nuclear and external regions.
This complex shape could be the result of the superposition of two effects:
overabundance of N
(mainly, in the nuclear starburst regions, where the stellar-wind of massive and
Wolf Rayet stars contribute to generate overabundance of N) and  shock heating
(in the external regions, according to the clear/clean
result obtained from the [S {\sc ii}]/H$\alpha$ map).

It is important to note, that
in the ``central regions", the [N {\sc ii}]/H$\alpha$ ratio
and FWHM(H$\alpha$) maps show a clear spatial correlation, i.e., the extended areas
of high values -in each map- show similar location.
Furthermore, these areas display similar elongated
shape and they are inside of the region of the bipolar outflow
(and close to the direction of the axis of the outflow cone).
These results are in good agreement with place where the ionization is
produced by shock heating, which could ionize the gas and broaden the
emission lines (see \S 4.5, Storchi-Bergmann, Wilson, \& Baldwin 1996 and
Lipari et al. 1997, for discussions related to the role of shocks processes
in the origin of high values of the [N {\sc ii}]/H$\alpha$ ratio).

\item
There are two strong peaks in the FWHM(H$\alpha$) map, which are associated
clearly to the main knots in the arm III (region 7) and to the area of very high 
negative velocity values. In addition, this map shows peaks (in the FWHM)
at the position of the main optical nucleus, and also in several central
and outer regions (see Fig. 5d).  These increase in the FWHM(H$\alpha$),
could be explained by the presence of multiple components inside of the
main/strong component (with different velocities) plus shock processes.

\end{enumerate}

Therefore, these results suggest that the {\it ``dominant" source of
ionization}, detected in the ``main" emission line component are:
(i) giant H{\sc ii} regions, in the nuclear and central area
(which are located in the nuclei and the spiral arms); and
(ii) the shock heating, in the external regions where we also found radial
filaments (Fig. 1b) and extended outflow (Fig. 2f); and where Scarrott et al.
(1996) detected the more extended polarization field known (which could be
associated mainly to the outflow/galactic--wind component). In addition,
for the central region, we also detected  composite ionization source:
H {\sc ii} regions plus shocks.

\vspace{5mm}

\centerline{{\it  3.3.3 Properties of the H{\sc ii} Regions and the Star 
Formation}}

\vspace{5mm}

We use in this section mainly the results of the relatively high resolution
spectrophotometry and kinematical data. 
The obtained emission line ratios of [S {\sc ii}]$\lambda\lambda$6731/6717
(Figs. 4 and 5) show a continuum variation in a range 
1.10--0.95, in the ionized gas. 
The highest value of the [S {\sc ii}]$\lambda\lambda$6731/6717
 ratio (i.e., the  highest electron
density condition, N$_{e}$) is reached in the main optical nucleus.
It interesting to note the good agreement between these high resolution
data and those present in Tables 4 an 5 (which were obtained with moderate 
resolution and for six strong knots). 

The corresponding range of N$_{e}$ obtained from the values
of [S {\sc ii}] $\lambda\lambda$6717, 6731 ratio and the relation given by
Osterbrock (1989) and Aller (1984) is $\sim$100--1000 cm$^{-3}$. 
The radial dependence of the [S {\sc ii}] ratio 
and the density could be explained by the fact that in the nuclear regions
there is a highest concentration of ionized gas. 
 
In low ionization giant H{\sc ii} regions and starburst  
(i.e., high metallicity giant H {\sc ii} regions) the determination of 
electron temperature in the ionized gas (T$_{e}$) and abundance is not 
straightforward, mainly by the absence of the lines [O {\sc iii}]$\lambda$ 4363
and [N {\sc ii}]$\lambda$ 5755. Therefore, the determination of the physical 
conditions (specially abundance) require a detailed study in a wide spectral
range $\lambda$3600 to 10000 \AA (see Pagel et al. 1979; Diaz et al. 1987, 
1990; Vilches et al. 1988; Terlevich 1993; Diaz \& Perez-Montero 1999). 
A range of electron temperature (T$_{e}$) in the ionized gas was obtained of
6000--7000 $^{\circ}$K; using the diagrams of Pagel et al. (1979), Dopita \& Evans
(1986) and Aguero \& Lipari (1991). 

In general the values of density and temperature in the ionized gas obtained 
in this paper (for different regions of NGC 3256) show also a good agreement 
with those obtained previously for all the central region of this galaxy. 
In particular, Storchi Bergman et al. (1995) found for the circumnuclear 
region (10$''\times$20$''$, aperture) a value of N$_e$ = 204 cm$^{-3}$ and 
T$_{e}$  = 6162 $^{\circ}$K. And, Aguero \& Lipari (1991) found for the nuclear region 
(3$''\times$6$''$ aperture) a value of  N$_e$ = 600 cm$^{-3}$ and T$_{e}$  
= 5900 $^{\circ}$K. In addition, these results are in agreement with the upper limit
of N$_e$ $\sim$ 1400 cm$^{-3}$ reported from Carral et al. (1994) based in
their measurement of the [O {\sc iii}](88/52 $\mu$m) IR line ratio. 

There are different and detailed methods to calculate the physical conditions 
of the ionizing source in giant H {\sc ii} regions (from optical emission lines), 
mainly using photoionization codes plus starburst model (see 
Garcia Vargas et al. 1997; Garcia Vargas \& Diaz 1994; Garcia Vargas, Bressan, 
\& Diaz 1995a, 1995b; Leitherer \& Heckman 1995). First,
we measured the temperature of the ionizing source using the optical emission 
lines ratios (from Table 5) and the results of photoionization and 
H{\sc ii} 
region/star--cluster models published by Pastoriza et al. (1993) for NGC 3310. 
And, we obtained in all their optical and near-IR diagrams (for the 6 regions)
a value of temperature of the ionizing source of
T$_{eff}$  $\sim$35.000 $^{\circ}$K (Figs. 6a,b,c,d; Pastoriza 
et al. 1993), for a parameter of ionization log(U) = -3.0 and solar abundance. 
In particular, using abundance-depend diagrams: 
[N {\sc ii}]$\lambda$6584/H$\alpha$ vs. 
([O {\sc ii}]$\lambda$3727+[O {\sc iii}]$\lambda$5007)/H$\beta$, and 
([S {\sc ii}]$\lambda\lambda$6717+6731 +[S {\sc iii}]$\lambda$9532)/H$\beta$ 
vs. ([O {\sc ii}]$\lambda$3727+[O {\sc iii}]$\lambda$5007)/H$\beta$
(their Figs 6c,d), we found  metallicity close to solar abundance.

These results are in excellent agreement with those obtained by
Joseph (1991), Doyon, Joseph, \& Wright (1994) from measurement of IR emission 
lines (He {\sc i}-2.06$\mu$m, Br$\gamma$-2.17$\mu$m, and H$_2$)
and IR absorption lines (CO band longward of 2.3 $\mu$m, Na {\sc i}
doublet--2.206$\mu$m, and Ca {\sc i} triplet--2.263 $\mu$m).
They found for the ionizing radiation field in this powerful starburst: 
T$_{eff}$  $\sim$35.000 $^{\circ}$K, lower mass cut-off 
$\sim$ 3--6 M$_{\odot}$, upper mass cut-off $\sim$ 30 M$_{\odot}$ and age 
12--27$\times$10$^6$ yr (using a deduced Initial Mass Function -IMF- 
slope of ${\alpha} \leq$ 2.2; where ${\alpha}$ come from
 dN(m)/dm $\sim$ m$^{-\alpha}$).
They used the starburst models of Telesco \& Gatley (1984) and Prestwich,
Joseph \& Wrigth (1994). In addition,
Moorwood \& Oliva (1994) found practically the same parameters/properties 
for the starburst in NGC 3256, using also near-IR emission lines data
([Fe {\sc ii}] 1.64$\mu$m, H$_2$ 2.12$\mu$m, Br$\gamma$ 2.17$\mu$m);
in particular, they obtained a lower mass cut-off 
$\sim$ 6 M$_{\odot}$, an upper mass cut-off $\sim$ 30 M$_{\odot}$, and IMF
slope $\alpha$ $\sim$2.5.

Recently,
Rigopoulou et al. (1996), and Lutz et al. (1996) studied the starburst of
NGC 3256, using  new ISO-SWS\altaffilmark{11} near and mid IR emission lines observations
([Ne {\sc iii}]/[Ne {\sc ii}] (15/12 $\mu$m), [Ar {\sc iii}]/[Ar {\sc ii}] 
(8.9/6.9 $\mu$m), [S {\sc iv}]/[S {\sc iii}] (10.5/18.7 $\mu$m)), 
plus new Sellmaier et al. (1996), 
Kurucz (1992), Kobo \& Sternberg (1996) stellar atmosphere and starburst 
models. They found for the ionizing radiation field (in the starburst/H {\sc ii} 
regions): T$_{eff}$  $\sim$41.000 $\pm$3000 $^{\circ}$K, upper mass cut-off 
$\sim$ 50--100 M$_{\odot}$ and age 10--20$\times$10$^6$ yr
(for log(U) = -2.5, solar abundance and Salpeter IMF with slope of 2.4). 
They suggest: {\it that the formation of very massive stars is probably not 
inhibited in NGC 3256}.
In \S 4.4 and Lipari et al. (1999) we analyze  this interesting result.

\altaffiltext{11}{ISO is the Infrared Space Observatory of the European Space 
Agency (ESA); and SWS is the ISO-instrument Short Wavelength Spectrometer}

The abundance were calculated, following the empirical method of Pagel et al. 
(1979) and Diaz \& Perez-Montero (1998, 1999) for low excitation H{\sc ii} 
regions, and we derived for the region 1+2:
O$^{+}$/H$^{+}$=8.7, O$^{++}$/H$^{+}$=8.1, N$^{+}$/H$^{+}$=7.9, 
S$^{+}$/H$^{+}$=6.7, O/H=8.8, S/H=6.6.
These are similar values to those obtained by Aguero \& Lipari (1991) and 
 Storchi Bergman et al. (1995) for the circumnuclear region. Therefore,
the higher ratio [N {\sc ii}]$\lambda$6584/H$\alpha$ observed in the nuclear 
region is probably due to metallicity plus shocks excitation.

\vspace{5mm}
\subsection{Multiwavelenght Continuum of NGC 3256}
\vspace{5mm}

The optical and near-IR continuum data of NGC 3256 obtained in this work
(\S\S 3.1 and 3.3) for the region 1+2 were compared to those extracted from 
the NED\altaffilmark{12} database (which were mainly obtained with wide 
aperture and therefore for the central region of the merger as a whole). 
We found a good agreement between these two set of data, and we plot all 
observations in Fig. 6 (log[Flux] vs. log[Frec]).
The data for this nearby galaxy cover from x-ray to radio wavelength,
and show clearly that:
(i) the shape of the continuum --of NGC 3256-- is very similar to those of 
galaxies with ``young" starbursts (e.g., M~82, NGC 7552; see Feinstein et al. 
1990);
(ii) the maximum flux is reached in the far IR  (in agreement with the measure 
of the ratio L$_{IR}$/L$_{B}$ $\sim$50); and
(iii) there are relatively strong flux emissions at radio and x-ray wavelength.
Therefore,
it is important to consider -in detail- the properties of the continuum 
in the main regions of the spectral range.

\altaffiltext{12}{NED is the NASA Extragalactic Database; NED is operated by the
Jet Propulsion Laboratory, California Inst. of Technology, under contract
with NASA.} 
 
At X-ray wavelength NGC 3256 is the only H{\sc ii} galaxy in the Boller et al. 
(1992) complete sample catalog of IRAS and ROSAT sources confirmed as 
luminous X-ray and IR emitter,  by Moran, Halpern \& Helfan (1994, 1996):
with L$_X   \geq$ 10$^{42}$ erg s$^{-1}$.
And, they found from ROSAT soft X-ray observations that L$_{X[total]} \sim$ 
2 $\times$ 10$^{42}$ erg s$^{-1}$, and from ASCA broad-band spectra that the 
L$_{X[0.3-10 kev]}$ = 4 $\times$ 10$^{41}$ erg s$^{-1}$. 
These values are greater than the highest X-ray luminosities known 
for H{\sc ii} galaxies (i.e., $\sim$ 10$^{41}$ erg s$^{-1}$, Fabbiano, Kim, \& 
Trinchieri 1992).
According to the main results presented in \S 3, i.e., the presence of an 
extended starburst with galactic--winds, the sources of the high X-ray 
emission could be attributed mainly to 
SN remnant plus emission from the hot phase of the ISM 
associated to the galactic--winds (see Suchkov et al. 1994). 
Furthermore, we measure a value of log(L$_{X}$/L$_{IR}$) = --2.8, which is a 
typical value for starburst galaxy with galactic--wind: 
M~82, NGC 253, NGC 3079, Arp 220, NGC 3628 and NGC 3690 (see HAM90: 
Table 7; and Ulrich 1978; Rieke \& Low 1972; Filippenko \& Sargent 1992; 
Lipari et al. 1994; Veilleux et al. 1994).

On the other hand, at UV wavelengths the continuum of NGC 3256
shows the ``reddest" slope with $\beta$ =+0.32 (for F$_{\lambda} \sim 
\lambda^{\beta}$; and $\alpha$ = --2 --$\beta$ = --2.32 for F$_{\nu} \sim 
\nu^{\alpha}$) in the UV spectra catalog of starburst galaxies (Kinney et 
al. 1993; Calzetti, Kinney \& Storchi-Bergmann 1994).
This slope is probably due to dust, and it is similar to that obtained for 
NGC 7552 ($\beta$ =+0.26), which presents several starburst processes 
(Dottori et al. 1986). The low value of EqW(H$\beta$) --in NGC 3256-- 
also suggest various processes of starburst or ionization by a small star cluster 
(see Terlevich, Terlevich, \& Franco 1993). 

At IR wavelength Graham et al. (1984), IRAS, Zenner \& Lenzen (1993), 
Moorwood \& Oliva (1994), Kotilainen et al. (1995), ISO, Boker et al. (1997)
observed the merger at near, mid and far--IR broad band continuum: 
I (0.82 $\mu$m), J (1.25 $\mu$m), H (1.65 $\mu$m), K (2.2 $\mu$m), 
L$'$ (3.75 $\mu$m), 12 $\mu$m, 25 $\mu$m, 60 $\mu$m, 100 $\mu$m.
They found mainly extended IR emission similar to our optical results 
(see \S 3.1 and Kotilanen et al. 1995), and SN rate of $\sim$2 yr$^{-1}$.
Therefore, probably the more important result obtained from the continuum 
of NGC 3256 is that the high IR luminosity come mainly from the
starburst component (i.e., the IR emission is ``extended", and there is not
evidence of AGN).

At radio wavelength Smith \& Kassin (1993) and Norris \& Forbes (1995)
observed NGC~3256 at 4.79 GHz (6 cm), 7.36 GHz (4 cm), 8.64 GHz (3 cm). 
They found extended diffuse radio emission 
plus several knots (located at the position of regions 1, 2 and 3, 4,
7, 10). Which have spectral index
$\alpha \sim$--1.0 (F$_{\nu} \sim \nu^{\alpha}$) 
typical for sycotron radiation generated by SN remnants
(and also typical of mergers/interacting galaxies; Smith \& Kassim 1993).
They obtained L$_{6 cm}$ = 5.1 $\times$ 10$^{39}$ erg s$^{-1}$ and SN 
rate of $\sim$3.8 yr$^{-1}$.

Therefore,
at different wavelength -from X-ray to radio-- the continuum and line
emission of NGC~3256 (obtained using mainly wide aperture, where the 
starburst is the dominant contribution) show an excellent agreement with 
the properties of an extended  {\it starburst with galactic--wind}
(see also \S 3.3). Furthermore, the radio spectral indices
between 3 and 6 cm., for the two main nuclei (region 1 and 3), are $\alpha =$ 
--0.78 and --0.86, which are consistent {\it with values of starburst 
nuclei} (Kotilainen et al. 1995).

\subsection{NGC 3256 Group of Galaxies}
\vspace{5mm}

It is important to study the properties of this galaxy in relation to 
those of the other members of the NGC~3256 group of galaxies, 
mainly for the following reasons: 
(i) NGC 3256C, the closer galaxy to NGC 3256 (at $\sim$150 kpc and 
$\sim$200 km s$^{-1}$), it is also an interacting/merger object, 
and these two systems 
plus the other members of the group could evolve to  mergers between mergers;
(ii) in this group (and also in the close NGC 3263 group) 
practically all the members show interacting/merger morphology (e.g., NGC 3256A
shows extended tidal tail);
(iii) the number of members, in NGC 3256 group, is not clearly defined; 
in particular two  
works based in the same sample of galaxies (Fouque et al. 1992; Garcia 1993) 
obtained different results, using similar computational algorithm. 
Also, Mould et al. (1991) study the probable presence of a cluster of galaxies
around the NGC 3256 group; and they found a high number of members for this 
cluster. In addition, the ESO NTT--EMMI images show clearly in 
the southern part of NGC 3256 two dwarf galaxies (Fig. 1b)
at a projected distance of $\sim$100--120$''$, from the nucleus. And, 
recently similar dwarf galaxies were found close to Arp 220 by Ohyama et al. 
(1999).

In particular,
Fouque et al. (1992)  using revised hierarchical algorithm  from 5554 galaxies
of Paturel et al. (1989a,b; the Lyon-Meudon Extragalactic Database 
catalogue), found 
4 members for this group: NGC 3256, NGC 3256C, NGC 3261 and ESO 317--G--20. 
Garcia et al. (1993) using also hierarchical algorithm, percolation 
and 2D method from 6392 galaxies of Paturel et al. (1989a,b, 1987), 
found three members: NGC 3256, NGC 3256C and NGC 3261. On the other hand,
Mould et al. (1991) study the probable presence of a cluster of galaxies
around the NGC 3256 group, in a radius of 4$^{\circ}$. And,
they found 10 members for this cluster: NGC 3256, NGC 3256A, 
NGC 3256B, NGC 3256C, NGC 3366, NGC 3318, NGC 3318B, ESO 263--G--51, ESO 
263--G--31, ESO 318--G--03 (and probably ESO 263--G--22); however, 
the members of the close NGC 3263 group are not included.

We are working in a detailed study (i.e., images, spectra, and kinematics) of the 
individual members and members-candidates  of the NGC 3256 and 3263 groups 
of galaxies, in a radius of 1$^{\circ}$. In particular,   
we started with the study of the galaxies NGC 3256C, NGC 3256A, NGC 3256B, 
NGC 3261, NGC 3366, NGC 3263/62 and ESO 317--G--20 (Lipari et al. 1999). 
In general, we found: (i) that NGC 3256C and NGC 3256A show 
merger morphology; specifically NGC 3256C shows very similar structure to 
the merger NGC 3310 (Pastoriza et al. 1993); 
(ii) the nuclear spectra of NGC 3256C shows interesting properties, with multiple
component systems; and 
(iii) NGC 3263 and 3262 show also interacting structure, plus a tidal tail. 

\newpage

\section{DISCUSSION}

In this section we discuss the main new results
obtained in \S 3. However, a detailed discussion
(including models for multiple merger, for asymmetrical spiral arms, 
and extended massive star formation process) 
will be published elsewhere (Lipari et al. 1999).

\subsection{Multiple Merger Model for NGC 3256 (and NGC 3256/3263 Groups)}
\vspace{5mm}

In \S 3.1, we found three independent asymmetrical spiral arms,
in the central 5 kpc region of NGC 3256. Our 3-dimensional
optical spectroscopy has shown that these arms are kinematically
independent. Furthermore, they appear to emanate from three different positions
which are identified as independent galactic nuclei (see \S\S 3.1.1 and 3.3).
Therefore, this fact strongly suggests that NGC 3256 is the result
of a collision between at least three galaxies, i.e., a multiple merger. 
Such multiple 
merger model has been recently proposed in order to explain the formation of 
ULIRGs such as Arp 220 (Taniguchi \& Shioya 1998; see also for general 
discussion about multiple mergers Barnes 1989; Weil \& Hernquist 1996). 

We consider -first- a not simultaneous  multiple merger model for NGC 3256.
Although the sense of rotation of arm I is the same as that of the arm III, the 
arm II has the opposite sense of rotation (with respect to those of the arms I 
and III). Therefore, it is expected that the precursor group of galaxies of NGC 
3256 consisted of three spiral galaxies one of which (a galaxy associated
with nucleus/region 2) has a retrograde orbit, or its orbit has been modified 
during the course of successive mergers.

NGC 3256 has a pair of long tails in both eastern and western sides (see
Fig. 1a). The formation of these two tidal tails requires that the final 
merging between two galaxies occur as a prograde merger (e.g., Barnes 
\& Hernquist 1996, 1992 and references therein).
However, one of this systems could be a previous merger between two objects.
Therefore one possible scenario is: (i) two galaxies associated with nucleus 1 
and 2, merge at first with a retrograde orbit; and then (ii) the merger remnant
(between the two galaxies) merges with the remaining galaxy -associated with 
nucleus 3- with a prograde orbit.
A more complete and detailed discussion of this scenario for NGC 
3256 will be presented by Lipari et al. (1999).

In relation to the three nuclei, it is important to remark that:
(i) the regions 1 and 3 were previously confirmed -at radio and mid-IR
wavelength- as two independent/main nuclei (by Norris \& Forbes 1995, Kotilainen et al.
1995, Zenner \& Lenzen 1993, and Moorwood \& Oliva 1994);
and, our VF shows that the kinematical center of NGC 3256
is located between these two regions; 
(ii) our data show that the region 2 presents very similar morphological,
kinematical and spectroscopic properties to those found for the region 1 (i.e.,
the main optical nucleus); specifically, in \S\S 3.1 and 3.3 we found that both
regions show  similar nuclear starburst  features (spectra, nuclear
spiral--disk, etc).

In addition, it is interesting  to consider the nature
 of the 3rd. source of polarization (Scarrott et al. 1996).
They suggested that probably it is a highly obscured nucleus, with a
strong starburst. We found an off-set of $\sim$1$''$, between the 3rd. source
of polarization and the nearest radio-peak (Norris \& Forbes 1995);
which is very weak and it is located at the same position of the optical
region 10. Therefore, the 3rd. source of illumination in the polarimetry
study could be associated to  H{\sc ii} regions located in the
region 10 (however our data do not exclude completely the presence of an obscured
nucleus, in this area).

It is important to note that -in \S 3.1- we found two small  nuclear
spiral--disks  (r $\sim$ 500 pc), around the regions 1 and 2 (the main
optical knots/nuclei), which were detected in the continuum and H$\alpha$
ionized gas emission (using HST-WFPC and ESO NTT-SUSI images). And,
the morphology suggest that these spiral--disk are probably counterrotating.
Sakamoto et al. (1999) found in Arp 220, two similar nuclear/small
conterrotating disk (r $\sim$100 pc) of molecular gas around the two optical
nuclei, with masses of $\sim$2.0 $\times 10^{9} M_{\odot}$.

The size of the tails in NGC 3256 can give us a first estimation 
of the age of the merger, i.e., since the last major merger began. 
If an escape velocity of 100 km s$^{-1}$ is assumed for the material in 
the tails (Colina et al. 1991) we obtain a value of 
$\sim$0.5 $\times$ 10$^{9}$ years. This result is in good agrement with the time
scale of mergers with prograde orbit: 0.5 $\times$ 10$^{9}$ yr (Barnes 1992;
Noguchi 1991).
It is important to note that the  starburst activity detected in this 
merger must have occurred very late in the history of the interaction,
since the age of the starburst is 5--25 $\times$ 10$^6$ yr (\S 3.3.3).
This is in agreement with the results published by Mihos \& Hernquinst
(1996) for models of starburst in disk/bulge/halo major mergers process
(their Figs. 4 and 2), where the extended massive star formation
occurred very late in the history of the interaction.

On the other hand, in the framework ot the multiple merger scenario
an alternative model is a  ``simultaneous" triple merger (and for this case
there are also several alternatives). 
A first simultaneous model could be between two gas rich massive spiral
galaxies (Sc), and a satellite galaxy (of one of these Sc systems). In order
to explore these type of scenarios it is required to perform detailed
numerical simulations, which are actually in progress (Lipari et al. 1999).
The firsts results of these kind of simulations
suggest that the two massive galaxies are the main component of the
merger event, and that the satellite generally merges later
(Taniguchi 1999, private communication). This type of composite model
-major plus minor mergers- could give an explanation of the very similar
properties (i.e., radio emission, mass, etc) found between the regions 1 and 3,
and some differences observed between these main/massive nuclei and
the nucleus/region 2.

Another interesting possibility is the simultaneous major
merger of three massive and gas-rich  Sc spirals; however this type of model
requires even more complex numerical simulations (Navarro 1999, private
communication).
Specially, for evolutive studies  of the induced massive star
formation process, in disk/bulge/halo major mergers
(see Mihos \& Hernquist 1996).

In addition, Weedman (1983), Rees (1984), Norman \& Scoville (1988), Taniguchi et al.
(1999), and others showed that in {\it nuclear starbursts} the presence of
star clusters with compact stellar remnants could form a supermassive black hole (SMBH)
with mass $\geq$ 10$^8$ M$_{\odot}$ (and/or large numbers of small accretors). 
This compact object, could evolve to an optically luminous quasars, and even to an 
elliptical/radio-galaxy (see Kormendy \& Sanders 1992; Terlevich \& Boyle 
1993; Lipari et al. 1994; Sanders \& Mirabel 1996). Recently,
the detection of extended starbursts --at scale of kpc-- in ULIRGs (Surace 
et al. 1998) generate new discussion related to this evolutive scenario. 
In NGC 3256, we found evidence of multiple merger and interesting conditions
in the starburst: extended massive star formation (r $\sim$ 5 kpc) ``plus" 
two/three compact nuclear--starbursts.
Therefore the observational results obtained in \S 3 combined with
theoretical studies suggest, for NGC 3256, that 
{\it ``the  multiple merger process"  (involving at least two/three compact
nuclear--starbursts) will probably evolve:}
(i) in each nucleus, to  SMBHs, which could be the final 
product of the merger of stellar compact remnants, or also
the product of the merger between two/three nuclei (see Tanguchi et al. 1999); and 
(ii) for all the system, to an ``elliptical host galaxy", 
when the multiple merger reaches the state of dynamic relaxation (as has been 
suggested by Graham et al. 1984). In \S 4.6, we discuss in more detail this
two points.

Finally, the presence of the interacting/merger system NGC 3256C very
close to NGC 3256 at 150 kpc and at $\sim$200 km s$^{-1}$ (plus the close 
presence of NGC 3263 Group of interacting galaxies and the merger
NGC 3256A; see \S 3.5)
suggest also the possibility --in the future-- of {\it mergers between 
mergers}. And, this fact reinforce the idea that multiple mergers (in 
ULIRGs and LIRGs),
in a scale of increase global mass, will likely generate elliptical
and even cD or radio galaxies (Schweizer 1982, 1990; Graham et al. 1984;
Sanders et al. 1988a; Lipari et al. 1994; Taniguchi \& Shioya 1998).
More specifically,  detailed photometric and kinematical studies
of ``typical" IR mergers (see Shier, Rieke \& Rieke 1994, 1996; and Shier
\& Fisher 1998, for references) show that: ``standard" mergers could
generate mainly low-luminosity ellipticals; and some ULIRGs could produce
high-luminosity and massive ellipticals. Our results -for NGC 3256- suggest that:
(i) multiple merger process combined with the induced extreme starburst/
galactic--wind (see \S 4.6 for details) are two main physical processes that
need to be considered in order to explain extreme properties in some LIRGs/ULIRGs;
and (ii) multiple merger process is a natural way to explain the formation
of some massive ellipticals from LIRGs/ULIRGs, specially in compact
groups and at the center of clusters of galaxies.

\subsection{The Origin and Properties of the Asymmetrical Spiral Arms}
\vspace{5mm}

Two interesting results obtained in this paper are: (i) the detection of
three asymmetrical spirals arms, in the nuclear and central regions; and (ii) 
that these arms show all {\it``very similar physical conditions", i.e.,
 high metalicity (low ionization) giant H {\sc ii} regions properties,
with temperature of the ionizing sources T$_{eff}$ $\sim$ 35000 $^{\circ}$K
and solar abundance, out to r $\sim$ 5 kpc!}.
In addition,  the two optical nuclei show also similar physical
properties (see \S\S 3.1 and 3.3).

The three asymmetrical arms have one-arm structure
(although arm I has a small counter arm, its length is much shorter).
In mergers process such asymmetrical --or one-arm-- spiral structures,
of young star formation and H {\sc ii} regions, were studied
previously, by Mihos \& Hernquist (1996) and Taniguchi \& Wada (1996).
Specifically, these arms may be driven mainly by:
(I) the triggering of young stellar component by the dynamical effect of
disk/bulge/halo major mergers process (Mihos \& Hernquist 1996: see their Figs.
4 and 2), where the inflow phase play a main role;
(II) the triggering of star formation by the effect of highly asymmetrical
gravitational potential, mainly in minor mergers (Taniguchi \& Wada 1996: see
their Fig. 7 where the model B$^+$ -for prograde orbit- shows very similar
properties to those detected in the arms and nuclear disks in NGC 3256).
 
The star formation with asymmetrical spiral arms structure may be driven by
a dynamical effect from a highly asymmetrical gravitational potential (case II),
from:
(i) binary (or multiple) black hole in a galactic nucleus,
(ii) two (or multiple) nuclear star clusters in a galactic nucleus,
(ii) nuclear star cluster plus a black hole in a galactic nucleus,
(iv) pair (or multiple) extragalactic nuclei.
And, with masses significantly different,
e.g., a factor 2 or more (Taniguchi \& Wada 1996).
For NGC 3256 we found (\S 3.1)  an elongated 
main structure in the very nucleus of region 1 (plus a compact knot;
see Fig. 1g); and
two compact knots (of different size) in the very nucleus of region 2 
(Fig. 1h). In addition, both regions have spectra with only 
nuclear starburst features (Figs. 3a, 3b and Tables 4 and 5). Therefore,
the presence -in both nuclei-  of two nuclear star clusters (or a 
bipolar elongate star cluster, or a star cluster plus a low mass nuclear
black hole) could be a likely option to consider, for the
origin of the asymmetrical spiral arms structures, as observed in NGC 3256. 
However, in the next section we consider that the processes of large infall
of matter plus outflow -in the ISM- are also important components for the
origin of these extended spiral arms and massive star formation (i.e., not
only dynamical effects).

In addition, more complex models are clearly required in order to consider
in detail multiple mergers process; where probably both major and minor
mergers processes are important for the origin of asymmetrical spiral arms of
young H {\sc ii} regions. And, detailed 2-D Fourier Analysis of these arms
(see Puerari \& Dottori 1997) and multiple merger simulations are actually in
progress in order to study the origin of these extended star formation
structures (Lipari et al. 1999).

For the obscured nucleus 3, it would be interesting to study its 
morphology at sub arc--second scale. This study could be made using 
mid-IR images with high resolution (similar to the PSF of the optical 
HST data: $\sim$0.1$''$) in order to detect 
similarities between the three main nuclear regions: e.g., the presence of
spiral disk. However, the near-IR HST-NICMOS images show that the arm III
is also connected or emanates from the region 3 (Fig. 1f).

On the other hand, the broad band BVI photometry (see Table 3) 
shows similar colors in the three/four asymmetrical spiral arms, and the three nuclei.
The arms show (B-V) color in the range $\sim$ 0.5--0.8 (Table 3) and all
show --at least-- one strong blue knot.
These properties were  also observed in the shape of the spectral continuum, 
in each individual region (Figs. 3).  
However, the presence of dust made these similar photometric properties less 
clear, specially in the arm 3.
The optical (B-V) colors observed (Table 3) are compatible with evolutive
models of star clusters of 10--15 $\times$ 10$^{6}$ yr
(Garcia-Vargas 1991: their Table 2.7 and Fig. 2.23); for a mass of the star 
cluster of M$_T$ = 3 $\times$ 10$^{8}$ M$_{\odot}$.

Morphological, spectroscopic, and photometric data show
similar properties in the arms and their nuclei (\S 3.1.1).
These are new results that probably require a detailed 
theoretical study.
However, a first and simple explanation is to consider that the physical 
conditions for the star formation process --in the central region--  
are similar at scale of r $\sim$5 kpc. 
The galactic--wind, detected in the central and nuclear regions of NGC 
3256 (\S 3.2.2, Fig. 2f, and Scarrott el al. 1996) could be the 
natural process in young starbursts (with a IMF that results in a high number 
of massive stars, SN event and the associated stellar wind; Robert et al. 1993)
that generated similar conditions (at scale of $\sim$5 kpc).

In addition, Taniguchi, Trentham \& Shioya (1998) and 
Shlosman \& Noguchi (1993) suggested new models for nuclear star formation
where the {\it infalling dense gas is unstable gravitationally and collapses 
to form massive gaseous clumps}, since these clumps are exposed to the 
external {\it high pressure that is driven by the galactic--wind},
they can collapse and then massive star formation may be induced in them.
Therefore, this is a first/possible explanation to this interesting new 
result (see for details \S\S\S 4.3, 4.4 and 4.5).

\subsection{The Complex Kinematics and Inflow in NGC 3256
(Ionized and Molecular Gas)}
\vspace{5mm}

Most of the feature of the H$\alpha$ VF could be reproduced by the superposition
of three rotation curves (RCs) extracted at three very different position angles.
They show the typical ``sinusoidal" shape of merger galaxies
(see \S 3.2.1).
We found that these RCs are defined mainly by the presence of outer
non-circular motions, since the decrease in the external parts
is higher than the values obtained by a Keplerian mass
decay. This seems to be also the case for sinusoidal RCs observed in similar
mergers: e.g., NGC 7252, 520, and 6240  (Schweizer 1982; Stanford \& Balcells 1990; 
Bland-Hawthorn, Wilson \& Tully 1991; Zepf 1993). And therefore, this
shape in the RCs is not directly related to the evolution of the halos and
the ``dark matter" of the original galaxies that collided.

An interesting test was performed using the molecular gas 
($^{12}$CO J=2-1) velocity map, published by Aalto et al. (1991,  their Fig. 6). 
First, we digitized this CO VF. 
This CO data have a resolution of $\sim$5-7$''$, and therefore in order
to compare this CO map to our H$\alpha$ data, we degraded 
our velocity field (using Gaussian filter) to match the CO resolution. 
The result was a smooth H$\alpha$ velocity field 
very similar to that obtained by Aalto et al. for CO (i.e.,
in each positive and negative part of the VF the two/three subregions were almost
combined in one extended lobe).
Therefore, the lack of resolution in distant mergers, could generate that 
the non-circular motion is considered as global rotation. In addition,
we extracted the RCs from the digitized CO and the ``smooth" H$\alpha$ velocity 
fields (for a PA$\sim$90$^{\circ}$ and for an aperture sector of 40$^{\circ}$). 
And, we found high amplitude residuals corresponding to each of the {\it
``sinusoidal/decreasing RCs"}; indicating that the smoothed VF does not represent
a real general motion. 

For the circumnuclear region, r $\leq$10$'' \sim$1.8 kpc (where we 
measured a general circular motion), we derive a Keplerian mass of
M$_{C.Nuclear}$ = 1.6$\pm$0.5 $\times 10^{10} M_{\odot}$. 
Since, the A$_{V} \sim$ 35 and L$_{IR}$/L$_{B} \sim$ 50 (Lutz et al. 1996), 
we used the IR-luminosity, in order to obtain the mass/luminosity ratio, and 
we derive a value of M$_{C.Nuclear}$/L$_{IR} \sim$ 0.05, for r $\sim$1.8 kpc. 
Previously, Feast \& Robertson (1978) obtained a lower limit for the total 
mass of $\sim$5 $\times 10^{10} M_{\odot}$ and M$_{Total}$/L$_{B} \sim$ 0.3. 
In addition, Wright et al. (1988) deduced a M$_{Nucl.Starburst}$ 
$\sim$7 $\times 10^{8} M_{\odot}$ and M$_{Nucl.Starburst}$/L$_{IR} 
\sim$ 0.01 
(from the kinematical values obtained by Feast \& Robertson 1978). 
These previous and the present results show a good agreement,
and also with those obtained for major mergers between gas-rich Sc 
spirals (Table 2), and mergers with strong nuclear massive 
star formation processes (see Wright et al. 1988).

Another interesting kinematical result, presented in \S 3.2, is the probable 
detection of inflow in a r $\sim$4$'' \sim$700 pc, close to the nuclear 
region (i.e., around the main optical nucleus; which is located near to the 
kinematical center: $\sim$400 pc). In particular, we found
-for NGC 3256- kinematical  and morphological evidences supporting the 
nuclear inflow scenario. And, these results are in agreement
to the predictions of numerical models for major and minor mergers: {\it ``where 
starbursts are driven by merger processes by depositing large amount of 
interstellar gas to the nuclear region"} (see  Mihos \& Hernquist 1996, 1994;
Taniguchi \& Wada 1996; Barnes \& Hernquist 1992). 

In particular, the detection of the superposed sinusoidal curves in the
nuclear region (r $\leq$ 4$''$), for a range of PA 40$^{\circ}$-130$^{\circ}$,
could be generated mainly by the presence of nuclear inflow or outflow
(Lipari et al. 1993b).
However, this PA's range (40$^{\circ}$-130$^{\circ}$) include the PA's range where
we found absence of the nuclear outflow component (45$^{\circ}$-80$^{\circ}$);
and therefore, the more likely explanation for this shape in the RCs
is the presence of nuclear inflow.
This is also in agreement with the observed near/far side orientation (\S 3.2).
Furthermore,
a very interesting fact is that the mean value --of the PA-- in the inflow region
(at PA $\sim$80$^{\circ}$) is practically perpendicular to the axis of the
bipolar outflow (at PA $\sim$150$^{\circ}$-160$^{\circ}$). Which is predicted
by models of outflow generated in nuclear star clusters (Perry 1992), and it
is expected if the inflow material (ionized and molecular gas) contribute to
the collimation process of the outflow material (Suchkov et al. 1996).

It is interesting to note, that in Arp 220,
all the ``luminous blue star cluster (LBSCs)" (Shaya et al. 1994)
are mainly located in the direction perpendicular to bipolar outflow's axis.
And we suggest, in \S\S 4.4 and 4.5, that in NGC 3256 the nuclear blue star
cluster was generated mainly by the interaction of the inflow and outflow
processes.

If we assume that the inflow rate is moderate/low (see Heller \& Shlosman
1994 for references), we can estimate the mass of 
the main nucleus, for a r $\leq$3$'' \sim$500 pc where we
detected a maximum in the inflow curve ($\Delta$V$_{max} \sim$  65 km s$^{-1}$).
Using spherical potential model (Bertola et al. 1991 and Lipari et al. 1993b)
and assuming that the inflow motion is close to circular:
we derived a mass of M$_{Region 1} \sim$  0.5 $\times 10^{9} M_{\odot}$.
For the region 3, a similar value of the mass could be assumed,
according to the result obtained in \S 3.2.1 (i.e., the kinematical center
is located close to the middle of the line that connect both nuclei).
These values of mass are in good agreement with those expected for galactic
nuclei with ``massive" stars clusters: $\sim$0.3 $\times 10^{9}
 M_{\odot}$ (Garcia Vargas 1991, and see \S 4.2); and also are similar to the
deduced value of the nuclear starburst mass 
$\sim$0.7 $\times 10^{9} M_{\odot}$ (Wright et al. 1988).

It is important to note,
that all detailed evolutive nuclear starburst models (see Tenorio-Tagle \&
Mu\~noz-Tu\~non 1997) required the presence of large infall of mater to the
central and nuclear regions (at $\sim$kpc scale),
as an indispensable condition in order to obtain large SFR
and starburst for 10$^{7}$ yr (as we observed in NGC 3256 and LIRG/ULIRGs).
These infalling material, which probably generates a giant spiral--disk
(Tenorio-Tagle \& Mu\~noz-Tu\~non 1997), could change strongly the conditions
in the ISM; and therefore play an important role in the evolution of the SFR and
the galactic--winds. In NGC 3256, probably the interaction
between the infalling material and the dynamical effect from an
asymmetrical gravitational potential, generate the massive star formation
process in the ``extended"  spiral arms and in the ``nuclear"
region.

The properties of the inflow detected in NGC 3256 (r $\sim$ 500 pc),
required to study also  models of small scale inflow ($\sim$10$^2$ pc).
Specifically, Perry (1992) and Williams \& Perry (1994)
found very interesting results modeling the outflow of a nuclear stellar
cluster and the flow in the ISM (using 2-D hydrodynamical, evolutionary
population synthesis and stellar evolutionary tracks models).
They found for the high mass loss stage (of the stellar cluster):
(i) super/hypersonic outflow with wide opening angle (close to spherical geometry);
(ii) half of the outflow-mass of the cluster is accreted;
(iii) this infall is roughly perpendicular to the axis of the outflow;
(iv) a peak luminosity L$_{Bol} \sim$ 10$^{47}$ erg s$^{-1}$ requires an
initial cluster stellar mass of M$_{Cluster} \sim$ 2 $\times$ 10$^9$ M$_{\odot}$
and initial average stellar density of $\sim$4 $\times$ 10$^8$ M$_{\odot}$ pc$^{-3}$.
The observed properties of the nuclear inflow detected in NGC 3256 show a good
agreement with these theoretical results.

\subsection{The Star Formation in NGC 3256 and LIRGs/ULIRGs}
\vspace{5mm}

The new results presented in \S 3, from the very nucleus to the end of the 
tidal tails of NGC 3256 (i.e., from r $\sim$15 pc to $\sim$40 kpc) are 
clearly compatible with an extended massive star formation plus a
 powerful galactic--wind, driven by the merger process.
And this extended massive star formation process is the main source of
high IR luminosity observed in this galaxy.
Similar results were found for practically all the prototype of 
LIRGs and ULIRGs (see Lutz et al. 1996; Genzel et al. 1988; Downes \& Solomon 
1998), where the evidences of the starburst and galactic--wind components are 
frequently less evident than in the nearby merger NGC 3256. 
In Arp 220, Smith et al. (1998) detected the presence of multiple radio SN 
in the nuclear regions, and they also found that the IR properties 
could be explained mainly by the presence of a strong starburst. 
Previously, Rieke et al. (1985), Heckman et al. (1987) and Lipari et al. 
(1994, 1993a) suggested a similar starburst/galactic--wind scenario for the luminous 
IR mergers Mrk 231, Arp 220 and NGC 6240; where the AGN component is probably 
strongly obscured or is not active. Recently, new ISO and CO
observations of luminous IR mergers, confirm that the starburst is mainly
the dominant source of IR luminosity in these systems (Genzel et al. 1998;
and Downes \& Solomon 1998).

The optical photometric, kinematic and spectroscopic results (\S 3)
complete --together with those obtained previously-- an
interesting multiwavelength set of data for NGC 3256, from UV to IR regions. 
These type of data are required in order to perform a complete study of the 
main properties of the star formation process (Scalo 1990). 
Previously, only for few starburst galaxies (e.g., NGC 7714) 
was possible to make this type of multiwavelength studies of star formation. 
Using this new set of data, we found interesting properties in 
NGC 3256:

\begin{enumerate}
\item
At optical wavelengths, 
the EqW(H$\alpha$+[N {\sc ii}]) for the region 1+2 is $\sim$430 \AA (Table 4). 
This value is one of the highest known for mergers 
(see Fig. 2 in Liu \& Kennicutt 1995: EqW(H$\alpha$+[N {\sc ii}]) for the
central region of 40 mergers). And this EqW
parameter is considered as a direct index of star formation rate (SFR), 
in starburst galaxies.
\item
At UV wavelengths, 
the measured values of EqW and blueshift in the two main absorption lines 
V$_{Si IV}$ =  --750 $\pm$60 km s$^{-1}$ and 
V$_{C IV}$  = --1070 $\pm$40 km s$^{-1}$
(Table 6), show a very interesting agreement with the
predictions of synthetic models of starburst with galactic--wind from massive 
stars. Specifically, our values are in agreement with
models of a instantaneous burst of 10 $\times$ 10$^{6}$ yr, Z = Z$_{\odot}$,
and M$_{u}$ = 60 M$_{\odot}$ (Robert et al. 1993: their Fig. 4).
\item
In the diagrams published by Scalo (1990): EqW[Si {\sc iv}]/EqW[C {\sc iv}]
vs. IMF Index, and Flux[O {\sc iii}]5007/FluxH$\beta$ vs. EqW(H$\beta$), our 
values show good agreement with models of starburst with a lower mass limit of 
$\sim$3 M$_{\odot}$, IMF index $\alpha$ = 2.5, and age of 10 $\times$ 10$^{6}$ 
yr; furthermore, our data fit well the starburst models for upper mass limit 
in a range of 30-80 M$_{\odot}$.
\item
Using the optical and near-IR emission line ratios of 
[S{\sc ii}]$\lambda$6717+6731/[S{\sc iii}]$\lambda$9069+9532   
(for region 1+2; Table 4) and the correlation between log(u) and 
[S{\sc ii}]/[S{\sc iii}] (from Diaz et al. 1991; and Garcia--Vargas 1991);
we found a deduced value of the ionization parameter log(u) = -3.0.
This is in good agreement with the value obtained in \S 3.3
(using the diagram/models of
Pastoriza et al. 1993; and for each individual regions). 
\item
At near-IR wavelengths, the absorption line of the Ca II triplet shows 
for the two prominent lines EqW($\lambda$8542+8662) $\sim$ 8.8 \AA (Table 6), 
which is a normal value for starburst galaxies (see Terlevich, Diaz, \& Terlevich 
1990, their Fig. 5). This value, is compatible to evolutive models of star 
clusters of 4--15 $\times$ 10$^{6}$ yr 
(Garcia-Vargas 1991: their Table 4.1 and Fig. 4.2); for standard parameters 
in the IMF. 

 And, therefore the range of age derived in this work for the extended
massive star formation detected in NGC 3256 is $\sim$5--25 $\times$ 10$^{6}$
yr (and the more probable value is: $\sim$10 $\times$ 10$^{6}$ yr). 

\end{enumerate}

An interesting point is the difference found, using ISO IR data, between
the properties of the starburst and the massive star formation
(i.e., T$_{eff}$ and the upper mass limit,
for the ionizing sources) obtained by Rigopoulou et al. (1996) and 
our results (\S 3.3.3).  
A first simple idea is that the use of optical observations,
could lead to biased low values for the high mass cutt-off, since the most
massive stars are not visible at optical wavelengths (Mirabel et al. 1998).
However, for NGC 3256, Joseph (1991), Doyon et al. (1994), Moorwood \& Oliva
(1994) using near-IR data obtained similar results to those presented in
\S 3.3.3. At least the differences in the T$_{eff}$ of the ionizing sources
could be easily explained by the fact that the T$_{eff}$ obtained by
Rigopoulou et al. (1996) is $\sim$41.000 $\pm$3000 $^{\circ}$K, but Lutz et
al. (1996) suggested that using better stellar atmosphere models than Kurucz
(1992) the T$_{eff}$ decreases in $\sim$2000 $^{\circ}$K.
Recently, there was also important improvement in modeling the ionizing
source of H {\sc ii} regions, using mainly the continuum from star clusters,
and not only from one type of early OB stars (Garcia-Vargas et al. 1997,
1995a,b; Pastoriza et al. 1993). Therefore, our value of T$_{eff}$ = 35.000  
$\pm$3000 $^{\circ}$K is the same --within the errors-- to that obtained by Rigopoulou 
et al. (1996). On the other hand, the results of the upper mass limit for the 
ionizing sources --which are close related to the T$_{eff}$--
are  model-dependent (Garcia-Vargas et al. 1997, 1995a,b);
and therefore, their discussion
required a specific study of the IMF, the star formation histories, 
the starburst process (i.e., constant SFR, 
$\delta$-burst, extended-burst, etc.; see Lipari et al. 1999). However, our
UV IUE data combined with syntetic starburst/galactic--wind models show
a good agreement with a upper mass limit of $\sim$60 M$_{\odot}$
(which is the same value obtained by Rigopoulou et al. 1996). 

Finally, an important result -obtained in \S 3.1- is the detection at the
center of the main optical nucleus (region 1),
of a blue elongate structure: probably a blue star cluster
of 63 pc $\times$ 30 pc; which we associate to the nuclear source of the 
starburst spectrum and the outflow.
This elongate structure, could be atributed to a proyection effect of a nuclear
stellar cluster with disk morphology (disk within disk); and Feldman et al.
(1982) suggested -from dynamical arguments-
this type of disk--structure for the very nucleus, in galaxies with nuclear
starbursts. In Arp 220,
at the center of the main/west nucleus, an elongate luminous blue star cluster
(of 75 pc) was also found by Saya et al (1994), using HST broad band images.
These two objects show very similar properties, and Taniguchi et al. (1998)
suggested that this type of ``luminous blue stars clusters"
(found mainly at the nuclear core of ULIRGs) are formed when
the nuclear infalling dense gas collapses to form massive gaseous clumps;
and then massive star formation is induced in them by the effect of
the external pressure that is driven by the galactic--wind. We detected,
in NGC 3256, all the components required for the formation of LBSCs
(in particular: inflow plus galactic--wind).  In the next section we also discuss
some properties of the extended massive star formation process.

\subsection{The Galactic-Wind in NGC 3256 and LIRGs/ULIRGs}
\vspace{5mm}

Graham et al. (1984), HAM90 and Scarrott et al. (1996) already suggested -for
NGC 3256- extended outflow and galactic--winds originated in a strong starburst.
However the results of the present work are the first direct kinematical
evidence for outflow, directly associated to the nuclear starburst: i. e.,
``galactic--winds".
Since, {\it there is not evidence of AGN properties},
in all the multiwavelength studies of NGC 3256 (see for detail Kotilainen et
al. 1995).

In \S 3.2, we found that the {\it ``nuclear" outflow} in NGC 3256 shows:
V$_{Nuc. OutFlow} \sim$ --350 km s$^{-1}$, FWHM $\sim$130 km s$^{-1}$,
bipolar structure (i.e.,
without component at PA 45$^{\circ}$--80$^{\circ}$), and ``very" wide opening
angle ($\theta$ $\sim$140$^{\circ}$).  In addition, we detected
that this blue component is extended in a r $\sim$5-6 kpc.
These results are in good agreement with the prediction made for 
the blowout phase of the galactic--wind, after 8 $\times$ 10$^{6}$ yr 
from the initial burst, i.e., in the beginning of the dusty starburst phase 
of type II SN (Suchkov et al. 1994; Lipari et al. 1994; Terlevich et al. 
1992; Norman \& Ikeuchi 1989). Furthermore,
the detection of ``radial filaments in practically all the
outer regions" of NGC 3256 (see \S 3.1.2 and Fig. 1b), plus the age of the
massive star formation process are also in good agreement
with the blowout phase of the galactic--wind. The polarimetry
study of NGC 3256, also shows the typical ``extended radial field", similar to
those found in galaxies in the blowout phase of the galactic--wind
(Scarrott et al. 1996).

In addition, we measured for the PA 110$^{\circ}$  the blue and red component
in the nuclear outflow: V$_{OF-blue}$ = --350$\pm$30 km s$^{-1}$ and
V$_{OF-red}$ = +350$\pm$50 km s$^{-1}$. In addition,
Fig 2f shows --for the PA 18$^{\circ}$-- similar values in the blue and red
outflow velocity, in the external regions.
Assuming the biconic surface model for the main origin of the outflow emission
lines (see for references HAM90; and Colina et al. 1991), similar values
of both outflow components suggest that the angle between 
the bicone axis  and the line of sight ($i_c$) is close to 90$^{\circ}$. And,
therefore using the value of the opening angle of the cone ($\theta$) we found a
value for the  velocity of the outflow V$_{OutFlow }$ = 370 km s$^{-1}$
(V$_{OF-blue}$ = --V$_{OutFlow}$$\times$cos[$i_c$ -$\theta$/2]). In this
simplest model the V$_{OutFlow}$, $\theta$, and $i_c$  are all constant and
independent of the distance from the nucleus.

Two complex and related aspects,
in the theoretical studies of the galactic--wind, are
the modeling of the collimation process and the evolution of the
geometry. The collimation process was studied mainly
for spiral galaxies; and for the evolution of the geometry (which appears
intimately connected with the structure of the ISM) two main ideas
were proposed: (i) bipolar galactic--wind arising in the plane parallel gaseous
atmosphere of a galactic disk (e.g., Tomisaka \& Ikeuchi 1988; Suchkov et al.
1994, 1996), which is important in the initial phases; and
(ii) spherical symmetric galactic--wind for the late phases (e.g., Mathews \&
Dones 1992; Chevalier an Clegg 1985), where the wind will lose
memory of the initial geometry.
Our results for NGC 3256 show composite structure:
bipolar outflow (with symmetry axis at the PA $\sim$150$^{\circ}$), but with
very wide opening angle ($\theta$ $\sim$140$^{\circ}$) which is practically 
a spherical geometry. This opening angle is higher than the mean values found
in the sample of HAM90: $\sim$60$^{\circ}$--80$^{\circ}$; but this range
was found mainly for spiral galaxies with galactic--wind (NGC 253, M82 and NGC
4945). However, our previous results for the ``superantennae"
(another IR merger) also show very wide opening angle in the outflow component:
$\theta$ $\sim$100$^{\circ}$ (Colina et al. 1991).
Therefore, the properties of the galactic--wind in mergers show clear
differences with those found in spiral galaxies.

In addition, we note that the orientation of the outflow bipolar axis
 is located close to the orientation of the photometric
and kinematic minor axis of the merger:
PA $\sim$170$^{\circ}$-160$^{\circ}$ and PA $\sim$00$^{\circ}$-170$^{\circ}$,
respectively (the PA of the photometric minor axis was obtained fitting
ellipses to the ESO NTT V broad band image).
Previously, Moorwood \& Oliva (1994) found a PA of $\sim$162$^{\circ}$ for
the near-IR photometric minor axis, fitting ellipses to the K$'$
image; and they show that over the central 15 kpc the K$'$ image is 
not yet that of a relaxed systems (r$^{1/4}$ profile).

In relation to the collimation process, it is important to note that the
observed small ionized gas spiral--disk (r $\sim$500 pc) and the star cluster 
found in the main optical nucleus of NGC 3256 (where the starburst
and the galactic--wind were generated; see \S 3.1.1),
are probably  associated to a small molecular disk. Similar to those detected
in Arp 220 (r $\sim$100 pc; Sakamoto et al. 1999) and this type of
small molecular disk could plays an important role in the initial collimation
process of the outflow (Suchkov et al. 1996).
Specifically, the resistence of the ISM -to the escape of the
outflow component- is lower in the direction perpendicular to the surface of
the molecular disk.

The kinematical shape observed for the {\it ``extended" outflow component} in the
emission lines [S {\sc ii}], [N {\sc ii}] and H$\alpha$ (Fig. 2f) 
is also in agreement with the results obtained for  models
of bipolar cone surface with very wide opening angle (see for example 
HAM90, their Fig. 19). Specifically, velocities with blue component will be
observed at both sides of the main nucleus (for each PA);
since this component came ``mainly" from the nearest part of the bicone
surface (which is centered on
the nuclear starburst). In addition, we also observed -for few regions- the 
red component of the extended outflow (Fig. 2f). We found similar
red component in the strong outflow/galactic--wind detected in the
south nucleus of the ultraluminous
IR merger IRAS 19254-7245 (the ``superantennae"; Colina et al. 1991).
In addition, the detection of extended radial filaments plus extended
polarimetry field support the idea that the outflow have also -at large
scale- a geometry close to spherical (very wide opening angle).

An important point to consider is that the outflow component shows high 
emission line ratios [N {\sc ii}]$\lambda$6584/H$\alpha$ and 
[S {\sc ii}]$\lambda$6717+6731/H$\alpha$ (see \S 3.2.2 and Fig. 2e). 
We found a similar result for the outflow component, in the
nearby starburst \& Seyfert nucleus of NGC 4945 (Lipari et al. 1997); and we 
suggested that this interesting result could be explained by high abundance 
of N generated by the presence of massive stars (specially, early WN type of 
Wolf Rayet stars) plus shock processes. In addition, we found in \S 3.3.2
(Figs. 5a-d) that also the ``main component" of the emission lines (of NGC 3256),
in the external regions with outflow, show very high values in these
two emission line ratios.

Using the emission line ratios obtained for the {\it nuclear outflow
component} (\S 3.2.2) and for the {\it external regions, in the main emission
line component} (\S 3.3.2), plus the diagnostic diagram of Veilleux \&
Oterbrock (1987) and HAM90 (Fig. 14), we found that
these values are inconsistent with gas heating by H{\sc ii} regions 
(where [N {\sc ii}]$\lambda$6584/H$\alpha$ $\leq$0.45, and
[S {\sc ii}]${\lambda 6717+31}$/H$\alpha$  $\leq$0.32, for H{\sc ii}
regions).
Furthermore, these observed emission line ratios are clearly consistent with 
shocks driven into clouds accelerated outward by a starburst with
galactic--wind: these values  -for the outflow and main components- are located
in the diagnostic diagram of HAM90 (their Fig. 14) in the region of 
SN remnants, shock heating, and Herbig--Haro objects
([S {\sc ii}]${\lambda 6717+31}$/H$\alpha$  $\geq$0.32; see Lipari et al.
1997). 

The simultaneous presence of ``galactic--wind" and  ``inflow" (in the main
optical nucleus of NGC 3256) could represent the first observational 
evidence that supports the two main arguments of the new models of nuclear
star formation (see \S 4.2 and Shlosman \& Noguchi 1993; Taniguchi, Trentham 
\& Shioya 1998), where the {\it infalling dense gas} collapses 
to form massive gaseous clumps; and since  these clumps are exposed to 
the  {\it high pressure that is driven by the galactic--wind},
they can collapse and then massive star formation may be induced in them.
Therefore, the {\it galactic-wind could be a natural process of a starburst 
driven by a starburst, in a cloudy protogalaxy that evolves by multiple mergers} 
(Berman \& Suchkov 1991). 
In this scenario the strong galactic--wind driven by the initial burst
of star formation plays an important role in the next step of the merger 
evolution; in particular for new star formation processes.
Finally, it is important to note that exactly in the same region where we
detected the galactic--wind (r$\sim$5-6 kpc), previously Sargent et al. (1989)
reported extended molecular H$_2$ gas (for r$\sim$6 kpc); and these are main
components in the extended star formation scenario proposed in this work. 

In addition, in order to study in detail the main and the feedbak processes 
that produce the extended massive star formation,
it is also important to consider another component (specially in systems with
strong starburst and galactic--wind): the {\it low-energy cosmic rays}, which
 -together with the shock process associated to the galactic--wind- play an important
 role as strong sources of heating and pressure of molecular clouds (see Allen 1992;
Suchkov, Allen \& Heckman 1993). These cosmic rays are  probably produced
at a rate proportional to SNs rate and could be evacuated by galactic--wind
(Suchkov et al. 1993).
And, studies of nuclear starburst/galactic--wind in IR mergers and spirals
galaxies have provided evidence for extremely high pressures in these regions:
three  to four order of magnitude higher than typical pressure in the ISM
in our Galaxy (Schaaf et al. 1988; HAM90).
Therefore, the contribution of cosmic ray could help to explain the high
pressure environment and the warm temperatures observed in the ISM of these
systems (and specially in the molecular clouds). These type of physical
conditions are required by our proposed scenario for the formation
of extended massive star formation in NGC 3256 (according to the general
scenario suggested by Shlosman \& Noguchi 1993, and Taniguchi et al. 1998).

Consequently, probably the interaction 
between dynamical effects, the galactic--wind (outflow), low-energy cosmic
rays, and the molecular+ionized gas 
(in the inflow phase) could be the possible mechanism that generate the 
{\it ``similar extended properties in the massive star formation, 
at scale of 5-6 kpc!"}.

Finally,
we found detailed kinematic, physical and morphological evidence --in NGC 
3256, Mrk 231, IRAS 19254-7245/super-antennae, and probably in IRAS 
07598+6508-- that {\it ``young starbursts" in luminous 
IR mergers} show the typical features of galactic--wind. In addition, we
detected in IRAS 22419-6049 the first type 1 IR--QSO that shows Wolf Rayet and
interacting features (Lipari \& Macchetto 1992;
previously Armus, Heckman \& Miley 1988 found Wolf Rayet features also in an 
IR--QSO, of type 2). Therefore, 
these and previous results for Arp 220, Mrk 266, Mrk 273, NGC 1222, NGC 1614, 
NGC 3690, NGC 4194, NGC 6240, and other objects (see HAM90, 
where 70$\%$ of their sample of galaxies showing galactic--wins are mergers!) 
strongly suggest that the relation between {\it merger, 
starburst, galactic--wind, and IR emission}, plays a main role in the evolution 
and formation of galaxies, SMBH, AGNs, QSOs, and consequently even in the 
evolution of some elliptical, cD and radio galaxies.

\subsection{The Relation Between: Merger, IR Emission, Starburst+Galactic-Wind, 
BAL+FeII QSOs, and Formation/Evolution of Galaxies}
\vspace{5mm}

In order to study merger, starburst and galactic--winds processes at high redshift,
 an interesting result obtained in this work is the fact that even for NGC 3256
 (the nearest advanced IR merger with  starburst and galactic--wind) 
{\it only a complete and detailed set of multiwavelength observations 
obtained using high and moderate spatial resolution  show  ``clean"  evidence 
of galactic--wind, nuclear-starburst and extended massive star formation} (which 
are superposed to merger features).
In particular, in order to study the ``real morphology" of the extended 
massive star formation,
it was required a spatial resolution better than 0.7$''$. And,
even the ESO NTT-EMMI data --obtained with 1.4$''$ resolution-- show mainly 
confused structures, which look like partial rings (Norris \& 
Fobes 1995 and Kotilainen et al. 1996, from near-IR and radio data obtained with
moderate spatial resolution, suggested that there are partial optical 
rings or tidal tails surrounding the central knots of NGC 3256).

In addition,  in order to detect the ``outflow component"  it was required
spectra obtained with moderate/high resolution (better than 100 km s$^{-1}$; see \S 3.2) 
and to cover a complete set of position angles (since this outflow component 
is weak or absent at the PA interval 45$^{\circ}$--80$^{\circ}$). 
In relation to this point it is important to note, that Moorwood \& Oliva 
(1994) suggested that --for NGC 3256-- there is not direct evidence of 
galactic--wind, in the spectroscopic data presented by HAM90. 
We also found that the kinematic presence of ``sinusoidal RCs (in mergers)" it 
is mainly indicative of non-circular motion and  is not a direct evidence of
merger process (Zepf 1993). Therefore, in order to obtain clear evidence of
starburst, galactic--wind, and mergers in distant galaxies/QSOs, it is required a 
detailed and complete multiwavelength set of high quality data (including: 
images, spectrophotometry, kinematics, etc.). And even, using this type of
data it is required a complete analysis of the different options.

The results presented in \S 3 are important, specially for the study of
IR QSOs/galaxies; and more specifically for the study of 
broad absorption line (BAL) IR selected QSOs. Mainly for the following reason: 
(i) Low et al. (1989), Boroson \& Meyer (1992) found that
IR selected QSOs show a 27$\%$ low-ionization BAL QSO fraction as compared
with 1.4$\%$ for the optically selected high-redshift QSOs sample (Weymann 
et al. 1991);
(ii) these objects are also {\it extreme/strong} Fe {\sc ii} emitters 
(Boroson \& Meyer 1992; Lipari 1994);
(iii) Lipari et al. (1994, 1993a); Lawrence et al.
(1996); and Terlevich, Lipari \& Sodre (1999)
proposed that the {\it extreme} IR+Fe {\sc ii}+BAL phenomena are related --at 
least in part-- to {\it the end phase of an ``extreme/strong starburst" and the 
associated ``powerful galactic--wind (GW)"}; and
(iv) {\it extreme} IR galaxies (LIRGs \& ULIRGs) are mainly mergers, see \S 1. 
At the end phase of a strong starburst, i.e., phase of type II SN
(8-60 $\times$ 10$^{6}$ yr from the initial burst;  
Terlevich et al. 1992; Norman \& Ikeuchi 1989;  Suchkov et al. 1994) 
 naturally appears giant galactic arcs (the galactic-bubble, 
and the broken shells),  
extreme Fe {\sc ii}+BAL systems and dust+IR-emission (Lipari 1994; Lipari et al.
1994, 1993a; Perry \& Dyson 1992; Dyson, Perry, \& Williams 1992; Perry 1992;
Scoville \& Norman 1996).

Specifically, in the starburst scenario two main models for the origin of
BAL systems were proposed:
(i) for IR dusty QSOs/galaxies, in the ouflowing gas+dust material the
presence of discrete trails of debris (shed by individuals
mass-loss stars) produce the BAL features (Scoville 1992; Scoville \&
Norman 1996); and
(ii) in SN ejecta, which is shock heated when a fast forward shock moves
out into the ISM (with a velocity roughly equal to the ejecta) and a reverse
shock moves and accelerating back towards the explosion center; and the
suppression of red-shifted absorption lines arise since SN debris moving
toward the central souce is slowed down much more rapidly -by the wind-
than is material moving away (Perry \& Dyson 1992; Perry 1992).
These two alternatives are probably complementary and both explain the main
observed properties of BAL phenomena (Scoville \& Norman 1996).

Recently, high resolution images of these IR selected BAL QSOs show in
practically ``all of these objects" the presence of arcs/shells
(Boyce et al. 1997; Stockton, Canalizo, \& Close 1998; Surace et al. 1998; 
Hines et al. 1999); and they are very similar
to those observed in Mrk 231 (the nearest merger+IR+GW+Fe {\sc ii}+BAL galaxy/QSO; 
see Lipari et al. 1994) and in Arp 220 (the nearest merger+IR+GW galaxy; see 
Heckman et al. 1987).
And, these ``circumnuclear and external arcs" could be associated mainly to 
the results of 
interaction of galaxies (tidal tails, ring, etc.) and/or to the final phase of 
the galactic--wind (i.e., the blowout phase of the galactic bubbles; see
Tomisaka \& Ikeuchi 1988; Norman \& Ikeuchi 1989; Suchkov et al. 1994). 
However, for distant AGNs and QSOs it is difficult to discriminate between 
these two related alternatives. 

Even for low redshifth BAL IR QSOs -like Mrk 231- there are different 
interpretations about the origin of these ``blue arcs".
In particular, Lipari et al. (1994) found clear evidence of a 
powerful/extreme nuclear starburst with galactic--wind in the circumnuclear 
region of Mrk 231, and we proposed a galactic--wind scenario for the origin 
of this blue arc. 
While  Armus, Surace et al. (1994) suggested the interaction between the main and an 
obscured/second nucleus, for the origin of this arc (furthermore, they also 
suggested that in this blue region and ``shell" there is not evidence of 
star-formation process).
Recently, HST--WFPC2(F439W) observations of Mrk 231 
confirm that this blue arc is a {\it ``dense shell of star-forming knots"} 
(see Surace et al. 1998: their Figs. 7 and the appendix). 
In addition, these HST--WFPC2 F439W broad-band images show in the circumnuclear 
region of Mrk 231 (r $\sim$ 1.5 kpc) blue spiral arms, similar to those 
observed in the central region of NGC 3256 (Lipari et al. 1999; Surace et 
al. 1998).
For the remaining selected IR QSOs, showing BAL+Fe {\sc ii} systems, the HST WFPC2 \&
NICMOS high resolution data suggest that the observed arcs/shells could be 
related mainly:
(i) in IRAS 04505+2958 and Mrk 231 to star-formation/outflowing material;
(ii) in IRAS 07598+6508 and IRAS 14026+4341 to interaction/merger process; and
(iii) in PG/IRAS 17002+5153 is --for us-- not clear, yet 
(see Lipari et al. 1999).

In the last years, Thompson, Hill \& Elston (1999)  and
Elston, Thompson, \&  Hill, (1994) reported more than 
15 QSOs at redshift $ 2 <$ z $< 5$, observed at the restwavelength of UV and 
optical Fe {\sc ii}+BAL spectral region. Aproximately  50$\%$ of these objects show 
``strong'' Fe {\sc ii} emission, and many of these objects are also BAL+IR QSOs.
In the starburst plus galactic--wind scenario  {\it extreme/strong Fe {\sc ii}+BAL+IR 
emitters are ``young IR QSOs/mergers"} where the starburst is 
probably the dominant source of output energy (Lipari et al. 1994, 1993a; 
Lipari 1994). Therefore, in order to study the real 
nature of these high redshift QSOs it is required a better underestanding of the 
merger, starburst, galactic--wind process in low redshift IR 
galaxies and QSOs (like NGC 3256, Arp 220, Mrk 231, NGC 3690,  
IRAS 07598+6508, IRAS 19254--7245, IRAS 22419--6049  and others).

In addition, Thompson et al. (1999) found a lack of iron abundance
evolution in high redshift QSOs: i.e., the absence of increase in the Fe {\sc ii}/Mg {\sc ii}
line ratio and Fe {\sc ii} equivalent width from the earliest epoch (z = 4.47 and
3.35) to the present.
Which represent a problem, since this fact indicate that 1 Gyr
may be an underestimate of the universe age at z = 4.47
(assuming that SN type Ia is the dominant source of Fe enrichment
in standard models of QSOs); and consequently q$_0$ could be
$\leq$ 0.20 for H$_0$ = 75 km s$^{-1}$ Mpc$^{-1}$. In our proposed starburst
scenario, both results: the detection of strong Fe {\sc ii} in QSOs at redshift
2 $\leq$ z $\leq$ 5 (or even at z $\geq$ 5),
plus the lack of iron abundance evolution in high redshift QSOs are in agreement
with the prediction of our models (Lipari et al. 1993a, 1994;
Lipari 1994; Terlevich et al. 1992, 1999). Specifically, by the reason that
in our scenario the time for the strong Fe enrichment in the shell
of SN type II and in the ISM is $\sim$ 8-60 $\times$10$^6$
yr (i.e., the end phase of an ``extreme starburst"). This time
is very short in relation to that required for the Fe enrichment
of the ISM by SN type Ia: i.e., 2 $\times$10$^9$ yr
(see Friaca \& Terlevich 1998). And, therefore --in our scenario-- the results
obtained by Thompson et al. (1999)  do not represent a problem
with the present idea of the age of the universe at z $\sim$ 5:
 $\sim$10$^{9}$yr for q$_0$ = 0.5 and H$_0$ = 75 km s$^{-1}$
Mpc$^{-1}$.

Very recently, detailed new-technology interferometric (IRAM and VLT) and
spectroscopic (ISO) studies with high resolution millimeter, near/mid-IR and 
radio data, confirmed the presence of {\it ``extreme" starburst} in ULIRGs: 
1000 times as many OB stars as 30 Dor in the IR mergers Mrk 231, Arp 
220, Arp 193 (see Downes \& Solomon 1998; Genzel et al. 1998 and Smith et al. 
1998, respectively).
In this paper, new-technology data (ESO NTT and HST) combined
with detailed and extensive optical observations (BALEGRE, CASLEO, CTIO) show new 
and clear evidences that NGC 3256 is another example of nearby luminous IR 
merger showing a {\it strong nuclear and extended massive star formation 
process}, with an associated {\it powerful galactic--wind}.

Consequently, the new detailed results presented in this work give support 
to the previous conclusion that {\it the properties of the merger and the 
associated ``extreme" starburst+galactic--wind} play a main role in the 
evolution of LIRGs/ULIRGs and  IR QSOs
(Rieke et al. 1985; Joseph \& Wrigth 1985; Heckman et al. 1987, HAM90;
Lipari et al. 1993a, 1994). More specifically, the results obtained for 
NGC 3256 are in good agreement with the following ideas: 
(i) ``multiple" merger process is very efficient way to generate 
extreme starburst and galactic--wind (and, associated dust+IR-emission)
 by depositing (inflow) the ISM gas in the central and nuclear regions; (ii) and 
then, play an important role the {\it interaction} between the merger and the 
extreme-starburst+galactic--wind process, and between the outflow and inflow  
material (which probably generate new star-formation processes, and 
{\it ``extreme" properties}). 

The {\it evolutive} end product of this interaction between 
mergers and extreme-starburst process could be: 
(i) SMBHs and IR-QSOs in the nuclear region, according to 
the conditions of the multiple-merger process, i.e., mainly the
nuclear compression of the ISM gas, the inflow accretion rate, etc
(Genzel et al. 1998; Downes \& Solomon 1998; Taniguchi et al. 1999); 
and (ii) elliptical, cD, radio galaxies for the multiple-merger 
process as a whole (Toomre 1977; Schweizer 1978, 1982; 
Barnes 1989; Barnes \& Hernquist 1992; Weil \& Hernquist 1996).

The results obtained in \S 3 are also interesting for the study 
of high redshift objects and {\it formation of galaxies}, since
it is expected that the properties of the {\it  initial collapse/merger}
and {\it starburst+galactic--wind} play -both- a main role, for practically
all the scenarios of galaxy formation  (see Larson 1974; Ostriker \&
Cowie 1981; Dekel \& Silk 1986; Ikeuchy \& Ostriker 1986; Lacey \& Silk 1991;
 Cole et al. 1994).
In particular, in luminous IR mergers the SFRs are close to 
those inferred of a galaxy forming itself: IR luminosities of $\sim$ 
10$^{11-12}$ L$_{\odot}$ implies SFR of $\sim$300--500 M$_{\odot}$ yr$^{-1}$,
if such SFRs are sustained for galaxy free-fall times  10$^{8-9}$ yr$^{-1}$, 
the total mass of newly formed stars would be 10$^{11}$  M$_{\odot}$ 
(see for detail HAM90). 
And therefore when we observed locally (in NGC 3256, Arp 220, Mrk 231, and 
others) the galactic--wind in luminous IR 
mergers we are probably observing the feedback processes from 
massive star formation that have important influence in determining the 
overall structure of galaxies in the {\it ``general dissipative collapse"} 
(see for details Ress \& Ostriker 1977; Silk 1977; Martin 1999; 
Norman \& Ikeuchy 1989; Bekki \& Shioya 1998;
HAM90; Tenorio-Tagle, Rozyczka, \& Bodenheimer 1990;  
Chevalier \& Clegg 1985).

More specifically, 
in the early stage of galaxy formation (when the SFR is expected to be
higher) the {\it ``galactic-wind"} play a decisive role in the 
feedback process: i.e., in the reheat of the ISM, which contributes to stop the 
initial collapse, and therefore to determine the overall structure of galaxies. 
Also the {\it ``galactic-wind"} plays a central role in  
some particular scenarios for galaxy formation, 
for example in the {\it ``explosive scenario"} (postulated by Ostriker \& Cowie 
1981) where the SNs explosions and galactic--wind are the process of 
self-regulation the SFR in young galaxies (see also McKee \& Ostriker 1977; 
HAM90; Lipari et al. 1994). And, these properties are very 
similar to those proposed --in \S\S 4.4 and 4.5-- for the extended massive star formation 
plus galactic--wind process, detected in NGC 3256 (for r $\leq$ 5 kpc).

SNs of type II  are highly concentrated in space and time, and formed from
massive stars (m$\geq$ 5 M$_{\odot}$) in young associations of tens/hundreds.
They are the main galactic objects capable to generate the blow-out phase
of the galactic--winds,
large amounts of dust+IR-emission, overabundance of Fe {\sc ii} (in the shell of 
SN-remnant and in the ISM), and also the BAL phenomena (Norman \& Ikeuchi 1989;
Perry \& Dyson 1992; Lipari et al. 1993a, 1994; Lipari 1994). However,
in the dusty nuclear regions of LIRGs and ULIRGs (with A$_{V}$ $\sim$10-1000 
mag; see Table 2 and Sakamoto et al. 1999; Lutz et al. 1996; Genzel et al. 1998), 
the presence of super/hyper--nova
could be detected only for nearby systems and using interferometric radio 
data (Smith et al. 1998). And therefore, for distant IR mergers (and objects
with composite source of nuclear energy: QSOs plus starburst)
a good indication of ``strong and extreme starbursts" could be 
features associate to the presence of a powerful galactic--wind 
(e.g., galactic-shell, bubbles, spectra with outflow components, etc).

\acknowledgments

The authors wish to thank J. Ahumada, J. Boulesteix, A. Diaz,
D. Garcia-Lambas, J. Melnick, A. Micol, A. Moorwood, J. Navarro, M. Nicotra,
S. d'Odorico, M. Pastoriza, T. Storchi-Bermann, A. Suchkov, E. Terlevich,
K. Thompson and  S. Zepf for discussions and help.
We would like to express our gratitude to the staff members and observing 
assistants at Bosque Alegre, CASLEO, CTIO and ESO Observatories. 
This work was based on observations made using the NASA and ESA HST and IUE
satellite, obtained from archive data at ESO-Garching and 
STScI-Baltimore. This work was made using the NASA Extragalactic Database NED;
which  is operated by the Jet Propulsion Laboratory, California Inst. of 
Technology, under contract with NASA. 
S. L. was supported in part by Grants from Conicet, Conicor, 
Secyt-UNC, and Fundaci\'on Antorchas (Argentina), and STScI (USA).
R. D. aknowledges a fellowship from FOMEC (FaMAF) at Cordoba University.

\newpage

\centerline{\bf REFERENCES}

\vspace{5mm}

\noindent
Aalto, S., et al. 1991, A\&A, 247, 291\\
Afanasiev, V., Dodonov, S., \& Carranza, G. 1994, Bol. Asoc. Argentina de 

Astron. 39, 160\\
Aguero, E., \& Lipari, S. L. 1991, Astrophy. \& Space Sc., 175, 253\\
Allen, R. 1992, ApJ, 399, 573 \\
Aller, L. 1984, Phyics of Thermal Gaseous Nebulae (Dordrecht: Reidel)\\
Armus, L., Heckman, T.M.,  \& Miley, G. 1988, ApJ, 326, L45 \\
Armus, L., Surace, J. et al. 1994, AJ, 108, 76 \\
Barnes, J. 1989, Nature, 338, 123\\
Barnes, J.  1992, ApJ, 393, 484 \\
Barnes, J., \& Hernquist, L. 1992, ARA\&A, 30, 705 \\
Barnes, J., \& Hernquist, L. 1996, ApJ, 471, 115\\
Bekki, K. \& Shioya, Y. 1998, ApJ, 497, 108 \\
Berman, B., \& Suchkov, A. 1991, Astrophy. \& Space Sc., 184, 169\\
Bertola, F., et al. 1991, ApJ, 373, 369\\
Bevington, P. 1969, Data Reduction and Error Analysis for the Physical

 Sciences, (New York: McGraw-Hill)\\
Bland,J., \& Tully, R. 1988, Nature, 334, 43\\
Bland, J., Taylor, K., \& Atherton, P. 1987, MNRAS, 228, 591\\
Bland-Hawthorn, J., Wilson, A. \& Tully R. 1991, ApJ, 371, L19\\
Boker, T., et al.  1997, PASP, 109, 827\\
Boller, T., et al.  1992, A\&A, 261, 57\\
Boroson, T., \& Meyer, K.  1992, ApJ, 397, 442\\
Boyce, P. J., et al.  1996, ApJ, 473, 760\\
Calzetti, D., Kinney, A., \& Storchi-Bergmann, T. 1994, ApJ, 429, 582\\
Carral, P., et al.  1994, ApJ, 423, 223 \\
Chevalier, R., \&  Clegg, A. 1985, Nature, 317, 44 \\
Clements, D., et al.  1996, MNRAS, 279, 459 \\
Cole, S., et al. 1994, MNRAS, 271, 181 \\
Colina, L., et al.  1997, ApJ, 484, L41 \\
Colina, L., Lipari, S. L., \& Macchetto, F.  1991, ApJ, 379, 113 \\
Condon, J. Anderson, M., \& Helow G. 1991a, ApJ, 376, 95 \\
Condon, J., Huang, Z., Yin, Q. \& Thuan, T.  1991b, ApJ, 378, 65 \\
Dekel, A. \& Silk, J. 1986, ApJ, 303, 39 \\ 
Diaz, A., et al. 1987, MNRAS, 226, 19\\ 
Diaz, A., et al. 1991, MNRAS, 253, 245\\ 
Diaz, A., \& Perez-Montero, E. 1998, ESO Workshop, in press\\
Diaz, A., \& Perez-Montero, E. 1999, MNRAS, submitted\\
Diaz, R., Carranza, G., Dottori, H. \& Goldes, G. 1999, ApJ, 512, 623\\ 
Djorgovski, S. 1994, in Proc. Conf. Mass-Transfer Induced Activity in 

Galaxies, ed.I. Shlosman (Cambridge: Cambridge University Press), 452\\
Dopita, M., \& Evans, G. 1986, ApJ, 307, 431\\
Doyon, R., Joseph, R. D., \& Wright, G. S. 1994, ApJ, 421, 101\\
Downes, D., \& Solomon, P. M. 1998, ApJ, 507, 615\\
Dyson, J., Perry, J., \& Williams, R. 1992, in Teesting the AGN Paradigm,

ed. S. Holt, S. Neff, \& M. Urry (New York: AIP), 548\\
Elston, R., Thompson, K., \& Hill, J. 1994, Nature, 367, 250 \\
Feast, M. \& Robertson, B. 1978, MNRAS, 185, 31 \\
Feintein, C. et al. 1990, A\&A, 239, 90 \\
Feldman, F., Weedman, D., Balzano, V., \& Ramsey, L. 1982, ApJ, 256, 427\\
Ferland, G.  1991, Internal Report 91-1, Astronomy Dept., Ohio State
University\\
Filippenko, A. \& Sargent, W. 1978, AJ, 103, 28 \\
Friaca, A., \& Terlevich, R. 1998, MNRAS, 298, 399\\
Garcia-Vargas, M., 1991, PhD Thesis, Univ. Atonome de Madrid, Spain\\
Garcia-Vargas, M., Bressan, A., \& Diaz, A. 1995a, A\&AS, 112, 13\\
Garcia-Vargas, M., Bressan, A., \& Diaz, A. 1995b, A\&AS, 112, 35\\
Garcia-Vargas, M., \& Diaz, A. 1994, ApJS, 91, 553\\
Garcia-Vargas, M., et al. 1997, ApJ, 478, 112\\
Genzel, R., et al. 1998, ApJ, 498, 579\\
Graham, G., Wright, G., Meikle, W., Joseph, R., \& Bode, M. 1984, Nature, 

310, 213\\
Hibbar, J. E.,  et al. 1994, AJ, 107, 67\\
Hines, D.,  et al. 1999, ApJ, 512, 140\\
Heckman, T.M., Armus, L., \& Miley, G. 1987, AJ, 93, 276\\
Heckman, T.M., Armus, L., \& Miley, G. 1990, ApJS, 74, 833 (HAM90)\\
Heller, C., \& Shlosman, I. 1994, ApJ, 424, 84 \\
Hutching, J. B.,  \& Neff, S. 1991, AJ, 101, 434\\
Ikeuchi, S. \&  Ostriker, J. 1988, ApJ, 301, 522 \\  
Joseph, R. D. 1991, in Massive Stars in Starbursts, eds.  C. Leitherer, 

N. Walborn, T.M. Heckman, C.A. Norman (Cambridge Univ. Press), 259\\
Joseph, R. D., \& Wright, G. S. 1985, MNRAS, 214, 87\\
Kinney, A., et al. 1993, ApJS, 86, 5\\
Kennicutt, R., Keel, W., \& Blaha, C. 1989, AJ, 97, 1022\\
Kormendy, J., \& Sanders, D. 1992, ApJ, 390, L53\\
Kotilainen, J., Moorwood, A., Ward, M., \&  Forbes, D.  1995, A\&A, 305, 107\\
Kobo, O. \& Sternberg,  A. 1996, in preparation\\
Kurucz, R. 1992, Rev. Mexicana Astron. Astrof., 23, 181\\
Lacey, C. \& Silk, J. 1991, ApJ, 381, 14 \\
Landolt, A. U. 1992, AJ, 104, 340 \\
Larson, R.  1974, MNRAS, 166, 585 \\
Laubert, A., et al. 1978, A\&AS, 35, 55 \\
Lawrence, A., et al.  1997, MNRAS, 285, 879 \\
Leitherer, C. 1991, in Massive Stars in Starbursts, eds. C. Leitherer, 

N. Walborn, T.M. Heckman, C.A. Norman (Cambridge Univ. Press), 1\\
Leitherer, C., \& Heckman, T. 1995, ApJS, 96, 9 \\
Lipari, S. L. 1994, ApJ, 436, 102 \\
Lipari, S. L., Colina, L., \& Macchetto, F.  1994, ApJ, 427, 174 \\
Lipari, S. L., et al. 1999, in preparation \\
Lipari, S. L., \& Macchetto, F.  1992, ApJ, 387, 522 \\
Lipari, S. L., Terlevich, R., \& Macchetto, F.  1993a, ApJ, 406, 451 \\
Lipari, S. L., Tsvetanov, Z., \& Macchetto, F.  1993b, ApJ, 405, 586 \\
Lipari, S. L., Tsvetanov, Z., \& Macchetto, F.  1997, ApJS, 111, 369 \\
Liu, Ch., \& Kennicutt Jr., R. 1995, ApJ, 450, 547\\
Low, F., Huchra, J., Kleinmann, S., \& Cutri, R.  1988, ApJ, 327, L41 \\
Low, F., Cutri, R., Kleinmann, S., \& Huchra, J.  1989, ApJ, 340, L1 \\
Lutz, D. 1991, A\&A, 245, 31\\
Lutz, D., et al. 1996, A\&A, 315, L137\\
Martin, C. 1999, ApJ, 513, 156 \\
Mathews, W. G., \& Dones J. 1992, Lick Observatory Bull., preprint \\
Mc-Carthy, P., Heckman, T., \& van Breugel, W. 1987, AJ, 93, 264\\
McKee, C.,  \& Ostriker, J. 1977, ApJ, 218, 148 \\
Melnick, J., \& Mirabel, I. F. 1990, A\&A, 231, L19\\
Mihos, C., \& Hernquist, L. 1994a, ApJ, 425, L13\\
Mihos, C., \& Hernquist, L. 1994b, ApJ, 431, L9\\
Mihos, C., \& Hernquist, L. 1996, ApJ, 464, 641\\
Mirabel, I. F., et al. 1998, A\&A, 333, L1\\
Mirabel, I. F., Dottori, H., \& Lutz, D. 1992, A\&A, 256, L19\\
Mirabel, I. F., Lutz, D., \& Maza, J. 1991, A\&A, 243, 367\\
Moran, E., Halpern, J., \&  Helfanf, D. 1994, ApJ, 433, L65\\
Moran, E., Halpern, J., \&  Helfanf, D. 1996, ApJS, 106, 341\\
Moorwood, A. F., \&  Oliva, E. 1994, ApJ, 429, 602\\
Mould, J. R., et al. 1991, ApJ, 383, 467\\
Noguchi, M. 1991, MNRAS, 251, 360 \\
Norman, C., \& Ikeuchi, S. 1989, ApJ, 395, 372\\
Norman, C., \& Scoville, N. 1988, ApJ, 332, 124\\
Norris, R., et al. 1990, ApJ, 359, 291\\
Norris, R., \&  Forbes, D. 1995, ApJ, 446, 594\\
Ohyama, Y., et al. 1999, ApJ, submitted (astro-ph/9903146) \\
Okumura S., et al. 1991, in IAU Symposium 146, Dynamics of Galaxies and 

their Molecular Distributions, ed. F. Combes (Dordretch: Kluwer), 425\\
Oliva, E., et al. 1995, A\&A, 301, 55\\
Osterbrock, D. 1989, in Astrophysics of Gaseous Nebulae and Active 

Galactic Nuclei, (University Science Book: Mill Valley)\\
Ostriker, J., \& Cowie, L. 1981, ApJ, 243, L127 \\
Pagel, B., et al. 1979, MNRAS, 189, 75\\
Paturel, G., et al. 1987, in Astronomy from Large Data Base, eds. Murtagh and 

Heck, ESO Workshop proc. 28, 435\\
Paturel, G., et al. 1989a, A\&AS, 80, 299 \\
Paturel, G., et al. 1989b, Catalogue of Principal Galaxies (Lyon: Obs. of Lyon

et Paris-Meudon)\\
Perry, J. 1992, in Relationships Between AGN and Starburst

Galaxies, ed. A. Filippenko (San Francisco: ASP Conf.S.31), 169\\
Perry, J., \& Dyson, R. 1992, in Teesting the AGN Paradigm, ed.

S. Holt, S. Neff, \& M. Urry (New York: AIP), 553\\
Phillips, A. 1993, AJ, 105, 486 \\
Plana, H., \& Boulesteix, J. 1996, A\&A, 307, 391 \\
Prestwich, A., Joseph, R., \& Wrigth, G. 1994, ApJ, 422, 73\\
Puerari, I.,  \& Dottori, H. 1997, ApJ, 476, L73\\
Ress, M., 1977, ARA\&A, 42, 471 \\
Ress, M.,  \& Ostriker, J. 1977, MNRAS, 179, 541 \\
Rieke, G., \& Low, F. 1972, ApJ, 176, L95 \\
Rieke, G., \& Low, F. 1975, ApJ, 197, 17\\
Rieke, G.,  et al. 1985, ApJ, 290, 116 \\
Rieke, G.,  et al. 1980, ApJ, 238, 24\\
Rigopoulou, D., et al. 1996, A\&A, 315, L125 \\
Robert, C., Leitherer, C., \& Heckman, T. 1993, ApJ, 418, 749 \\
Rubin, V., Graham, J., \& Kenney, J. 1992, ApJ, 394, L9\\
Rubin, V., \& Ford, W. 1983, ApJ, 271, 556\\
Sakamoto K., et al. 1999, ApJ, 514, 68 \\
Sanders, D. B., Egami, E., Lipari, S., Mirabel, I., \& Soifer B. T. 1995, 

AJ, 110, 1993 \\
Sanders, D. B., \& Mirabel, F. 1996, ARA\&A, 34, 749\\
Sanders, D. B., Scoville, N., \& Soifer B. T. 1991, ApJ, 370, 158 \\
Sanders, D.B., Soifer, B.T., Elias, J.H., Madore, B.F., Matthews, K.,

 Neugebauer, G., \& Scoville, N.Z. 1988a, ApJ, 325, 74\\
Sanders, D.B., Soifer, B.T., Elias, J.H., Neugebauer, G., \& Matthews, K.

 1988b, ApJ, 328, L35\\
Shaya, E., et al. 1994, AJ, 107, 1675\\
Sargent, A., Sanders, D.B., \& Phillips, T. 1989, ApJ, 346, L9\\
Scalo, J. 1990, in Windows of Galaxies, eds. Fabbiano et al.

(Dordrecht: Kluwer), 125 \\
Scarrott, S. M., Draper, P. W., \& Stockdale, D. P. 1996, MNRAS, 279, 1325\\
Scott, D.,  et al. 1998, Nature, 394, 219\\
Scoville, N. Z. 1992, in Relationships Between AGN and Starburst

Galaxies, ed. A. Filippenko (San Francisco: ASP Conf.S.31), 159\\
Scoville, N. Z., \& Norman, C. 1996, ApJ, 451, 510 \\
Scoville, N. Z., Sargent, A., Sanders, D., \& Soifer, B. 1991, ApJ, 366, L5 \\ 
Scoville, N. Z., \& Soifer, B.T. 1991, in Massive Stars in Starbursts, eds. C. 

Leitherer, N. Walborn, T.M. Heckman, C. Norman (Cambridge Univ. Press), 233\\  
Silk, J. 1977, ApJ, 211, 638\\
Schweizer, F.  1980, ApJ, 237, 303\\
Schweizer, F.  1982, ApJ, 252, 455\\
Schweizer, F.  1986, Science, 231, 227\\
Schweizer, F.  1990, in Dynamics and Interactions of Galaxies, 

ed. R. Wielen, (Springer--Verlag, Heidelberg), 60\\
Sekiguchi, K. \& Anderson, K. 1987, AJ, 94, 644 \\
Sellmaier, F., et al. 1996, A\&A, 305, L37\\ 
Sersic, J. L. 1959, Zs. Ap, 47, 9\\
Shaaf, R., et al. 1989, ApJ, 336, 762 \\
Shier, L.,  \& Fischer, J. 1998, ApJ, 497, 163 \\
Shier, L., Rieke, M., \& Rieke, G. 1994, ApJ, 433, L9 \\
Shier, L., Rieke, M., \& Rieke, G. 1996, ApJ, 470, 222 \\
Shlosman, I., \& Noguchi, M. 1993, ApJ, 414, 474 \\
Smith, E., \& Kassim, N. 1993, AJ, 105, 46\\
Smith, H., Lonsdale, C., Lonsdale, C., \& Diamond, P. 1998, ApJ, 493, L17\\
Soifer, B.T., et al. 1984, ApJ, 283, L1\\
Soifer, B.T., et al. 1987, ApJ, 320, 238\\
Stanford, S., \& Balcells, M. 1990, ApJ, 355, 59 \\
Stocton, A., Canalizo, G., \& Close, L. 1998, ApJ, 500, L121\\
Stone, R.,  \& Baldwin, J. 1983, MNRAS, 204, 357\\
Storchi-Bermann, T., Kinney, A., \& Challis, P. 1995, ApJS, 98, 103\\
Storchi-Bermann, T., Wilson, A., \& Baldwin, J. 1996, ApJ, 460, 252\\
Suchkov, A., Allen, R., \& Heckman, T. 1993, ApJ, 413, 542\\   
Suchkov, A., Balsara, D., Heckman, T., \& Leitherer, C. 1994, ApJ, 430, 511\\   
Suchkov, A., Berman, V., Heckman, T., \& Balsara, D 1996, ApJ, 463, 528\\   
Surace, J., et al. 1998, ApJ, 492, 116\\
Taniguchi, Y., Ikeuchi, S., \& Shioya, K. 1999, ApJ, 514, L9\\
Taniguchi, Y., \& Shioya, K. 1998, ApJ, 501, L67\\
Taniguchi, Y., Trentham, N., \& Shioya, K. 1998, ApJ, 504, L79\\
Taniguchi, Y., \& Wada, K. 1996, ApJ, 469, 581\\
Telesco, C., \& Gatley, I. 1984, ApJ, 284, 557\\
Tenorio-Tagle, G., \& Mu\~noz-Tu\~non, C. 1997, ApJ, 478, 134\\
Tenorio-Tagle, G., Rozyczka, M., \& Bodenheimer, P. 1990, A\&A, 237, 207\\
Terlevich, E. 1993, Rev. Mexicana Astron. Astrof., 27, 29\\
Terlevich, E. Diaz, A., \& Terlevich, R. 1990, MNRAS, 242, 271\\
Terlevich, E. Terlevich, R., \& Franco, J. 1993, Rev. Mexicana Astron. 

Astrof., 195\\
Terlevich, R., \& Boyle, B. 1993, MNRAS, 262, 491\\
Terlevich, R., et al. 1991, A\&AS, 91, 285\\
Terlevich, R., et al. 1992, MNRAS, 255, 713\\
Terlevich, R., Lipari, S., \& Sodre, L. 1999, MNRAS, in preparation\\
Thompson, K., Hill, J., \& Elston,  R. 1999, ApJ, 515, 487 \\  
Tomisaka, K., \& Ikeuchi, S. 1988, ApJ, 330, 695 \\  
Toomre, A.  1977, in The Evolution of Galaxies and Stellar Population 

(New Haven: Yale University Observatory), 401\\
Toomre, A., \& Toomre, J. 1972, ApJ, 178, 623\\
Ulrich, M. H. 1972, ApJ, 178, 113.\\
Ulrich, M. H. 1978, ApJ, 219, 424.\\
Veilleux, S., \& Osterbrock, D. E. 1987, ApJS, 63, 295\\
Veilleux, S., et al. 1994, ApJ, 433, 48\\
Veilleux, S., et al. 1996, ApJS, 98, 171\\
Veilleux, S., et al. 1999, ApJ, preprint astroph \\
Vilches, J., et al. 1988, MNRAS, 235, 633\\
de Vaucouleurs, G., \& de Vaucouleurs, A.  1961, Mem. RAS, 68, 69\\
Weedman, D. W. 1983, ApJ, 266, 479\\
Weedman, D., et al. 1981, ApJ, 248, 105\\
Weil, M. L., \& Hernquist, L. 1996, ApJ, 460, 101\\
Weymann, R., et al. 1991, ApJ , 373, 23\\
White, S. 1994, in Proc. Conf. Mass-Transfer Induced Activity in 

Galaxies, ed.I. Shlosman (Cambridge: Cambridge University Press), 464\\
Wright, G., Joseph, R. D., Robertson, N., et al. 1988, MNRAS, 233, 1\\ 
Williams, R., \& Perry, J. 1994, MNRAS, 269, 538 \\
Zenner, S., \& Lenzen, R. 1993, A\&AS, 101, 363 \\ 
Zepf, S. 1993, ApJ, 407, 448\\

$${
\begin{tabular}{lcccl}
\multicolumn{5}{c}{\bf Table 1: Journal of observations of NGC~3256}\\
\hline
\hline
Date           & Telescope/ & Spectral Region  & Exp. Time  & Comments \\
               & Instrument &                  &seconds     & \\
\hline

1990 March 03   & CTIO 1.0m  &$\lambda \lambda$3600-6900\AA & 1200 x 2 & 
PA 90$^{\circ}$ long-slit  \\

1993 Feb 22     &ESO EMMI 3.5m& V                        & 300 x 2 & images\\
1993 April 09   &ESO SUSI 3.5m& $\lambda \lambda$5013/139 \AA& 300&
[O {\sc iii}] redshifted \\
                &            & $\lambda \lambda$5111/55 \AA  & 300     & 
[O {\sc iii}] continuum \\
                &            & $\lambda \lambda$6571/115 \AA & 300     & 
H$\alpha$+[N {\sc ii}] redshifted\\
                &            & $\lambda \lambda$7018/65 \AA  & 300     & 
H$\alpha$+[N {\sc ii}] continuum \\

1994 Sept. 7    &HST/WFPC2   & F814W, $\lambda \lambda$7970/1531 \AA & 800 x 2 & \\
                &            & F450W, $\lambda \lambda$4521/958  \AA & 900 x 2 & \\
1997 Decemb. 5  &HST/NICMOS &F160W, $\lambda \lambda$1.60/0.20$\mu$m& 48& NIC2\\
                &           &F110M, $\lambda \lambda$1.10/0.10$\mu$m& 32& NIC1\\
                &           &F222M, $\lambda \lambda$2.22/0.07$\mu$m& 80& NIC2\\
                &           &F237M, $\lambda \lambda$2.37/0.07$\mu$m& 80& NIC2\\
                &           &F166N, $\lambda \lambda$1.66/0.01$\mu$m& 96& NIC1\\
                &           &F190N, $\lambda \lambda$1.90/0.01$\mu$m&128& NIC2\\
                &           &F215N, $\lambda \lambda$2.15/0.01$\mu$m&300& NIC2\\

1997 March 11   &CASLEO 2.15m&B               & 600 x 3 &  images        \\
                &            &V               & 300 x 3 &           \\
                &            &I               & 240 x 3 &           \\
1997 March 12   &            &$\lambda \lambda$4100-7500 \AA & 1200 x 2   & 
PA 90$^{\circ}$ long-slit  \\ 

1997 April 30 &B. ALEGRE 1.5m&$\lambda \lambda$6400-6900 \AA & 1800 x 4 & 
PA 80$^{\circ}$ long-slit  \\
1997 May 5     &  EMF        &$\lambda \lambda$6400-6900 & 2400 x 4 & 
PA 110$^{\circ}$, 80$^{\circ}$, 00$^{\circ}$   \\
1997 May 6     &             &$\lambda \lambda$6400-6900 & 1800 x 6 & 
PA 90$^{\circ}$, 00$^{\circ}$, 110$^{\circ}$ \\
1997 May 7     &             &$\lambda \lambda$6400-6900 & 1800 x 4 & 
PA 135$^{\circ}$, 45$^{\circ}$ \\
1997 May 8     &             &$\lambda \lambda$6400-6900 & 1800 x 4 & 
PA 25$^{\circ}$, 155$^{\circ}$ \\
\hline
\end{tabular}
}$$

\newpage

$${
\begin{tabular}{lcccl}
\multicolumn{5}{c}{\bf Table 1: Continuation}\\
\hline
\hline
Date           & Telescope/ & Spectral Region  & Exp. Time  & Comments \\
               & Instrument &                  &seconds     & \\
\hline

1997 May 9     &             &$\lambda \lambda$6400-6900 & 1800 x 4 & 
PA 65$^{\circ}$, 110$^{\circ}$ \\
1997 May 11    &             &$\lambda \lambda$6400-6900 & 1800 x 4 & 
PA 25$^{\circ}$, 125$^{\circ}$ \\
1997 May 12    &             &$\lambda \lambda$6400-6900 & 1800 x 2 & 
PA 45$^{\circ}$ \\
1997 June 6    &            &$\lambda \lambda$6400-6900  & 1800 x 4 & 
PA 56$^{\circ}$, 168$^{\circ}$ \\
1997 June 7   &             &$\lambda \lambda$6400-6900  & 2400 x 2 & 
PA 145$^{\circ}$ \\
              &             &$\lambda \lambda$4800-6900  & 1800 x 2 & 
PA 90$^{\circ}$ 7$''$ S \\
1997 June 8   &             &$\lambda \lambda$6400-6800  & 1800 x 2 &
PA 13$^{\circ}$ \\
              &             &$\lambda \lambda$4800-6900  & 1800 x 2 & 
PA 90$^{\circ}$ 7$''$ S \\
1997 June 9   &             &$\lambda \lambda$6400-6900  & 1800 x 2 & 
PA 100$^{\circ}$ \\
1997 July 4   &             &$\lambda \lambda$6400-6900  & 1800 x 2 & 
PA 40$^{\circ}$ \\
1997 July 5   &             &$\lambda \lambda$6400-6900  & 1800 x 4 & 
PA 130$^{\circ}$, 40$^{\circ}$ \\
1997 July 13  &             &$\lambda \lambda$6400-6900  & 2400 x 2 & 
PA 90$^{\circ}$ 7$''$S \\
1997 July 15  &             &$\lambda \lambda$6400-6900  & 1800 x 2 & 
PA 90$^{\circ}$ 7$''$S \\
1998 February 26&           &$\lambda \lambda$6400-6900, V, I  & 2400 x 2 & 
PA 90$^{\circ}$ (N3256C, N3261) \\
                &           &$\lambda \lambda$6400-6900, V, I  & 2400 x 2 & 
PA 108$^{\circ}$ (N3263) \\
1998 February 27&           &$\lambda \lambda$6400-6900  & 2700 x 4 & 
PA 18$^{\circ}$  \\
                &           &$\lambda \lambda$6400-6900  & 2700 x 4 & 
PA 82$^{\circ}$ Tail NE 48$''$N\\
1998 May 29     &             &$\lambda \lambda$6400-6900  & 1800 x 6 & 
PA 70$^{\circ}$  \\
1998 May 30     &             &$\lambda \lambda$6400-6900  & 2700 x 2 & 
PA 50$^{\circ}$  Tail NE 33$''$N\\
1999 Jan 14   &             &$\lambda \lambda$6400-6900, V, I  & 2700 x 2 & 
PA 90$^{\circ}$  (N3262)\\
1999 Feb 19  &             &$\lambda \lambda$6400-6900, V, I  & 2700 x 2 & 
PA 67$^{\circ}$  (N3256C)\\
1999 March 22   &             &$\lambda \lambda$6400-6900  & 2700 x 2 & 
PA 90$^{\circ}$, 137$^{\circ}$  (N3256A/3256B)\\
1999 March 23   &             &V and I                  & 600 x 24 & 
N3256A,B,C; N3261/62/63\\
\hline
\end{tabular}
}$$

\newpage

$${
\begin{tabular}{lcccccc}
\multicolumn{7}{c}{\bf Table 2: General Properties of NGC~3256 and Luminous IR 
Ongoing Mergers}\\
\\
\hline
\hline

       Parameters             &N4038/9&  N3256&  A220 &N6240&  N7252& Mk 231\\
\hline

                              &       &       &       &       &       &       \\
{\it Parameters Adopted:}     &       &       &       &       &       &       \\
D (Mpc, H$_{0}$= 75)          & 22    & 35    &  77   & 98    & 64    & 170   \\
Inclination                   & --- &40--45$^{o}$& --- & --- &41$^{o}$& ---   \\
                              &       &       &       &       &       &       \\
{\it Parameters Derived:}     &       &       &       &       &       &       \\
Systemic Velocity (Km s$^{-1}$)& 1650 & 2817  & 5450  & 7300  & 4800  & 12450 \\
M$_{B}$                       & -22.4 &  -22.6&(-20.5)&(-21.0)& -21.2 & -22.0\\
log L$_{IR}$ (L$_{\odot}$)    &  10.30&  11.48&  12.19& 11.82 & 10.36 &  12.53\\
L$_{IR}$/L$_{B}$              &   1   & 50    & 150   &   45  & 1     &  300  \\
SFR (M$_{\odot} yr^{-1}$)     &  ---  & 80    & 300   &  100  & ---&$\sim$1000\\
SN Rate (yr$^{-1}$)           &  ---  &  2    &  3    &  1.0  &  ---  & 5 \\
d$_{nucl.}$ (kpc, H$_{0}$= 75)&  6.4  &   1.0 &  0.4  &  0.5  &   0.0 &   0.0 \\
A$_{V nucl.}$ (mag.)          &  70   & 35   &100-1000&20-40 &$\sim$1&$\geq$20\\
                              &       &       &       &       &       &       \\
Dynam. Total Mass
(M$_{\odot} \times$10$^{10}$) &  8.0   &$\sim$5&$\sim$1& (8)  & 4.0 & ---  \\
Molec. H$_{2}$ Mass     
(M$_{\odot} \times$10$^{9}$)  & 5.0   & 30    &  60.3 & 20.0  & 3.6   &  58.9 \\
Atomic HI Mass     
(M$_{\odot} \times$10$^{9}$)  & 9.8   &  ---  &  --- &  --- & 3.4&$\leq$22.4\\
Ionized H{\sc ii} Mass     
(M$_{\odot} \times$10$^{7}$)  &  ---  &  ---  &  ---  & ---   & 11    &  ---  \\
Dust Mass     
(M$_{\odot} \times$10$^{7}$)  &  1.4  & (2.0) & 12.0   & 7.0 &$\leq$1.5&  12.0\\
                              &       &       &       &       &       &       \\
\hline

\end{tabular}
}$$

\newpage

$${
\begin{tabular}{lcccccccc}
\multicolumn{9}{c}{\bf Table 3: Optical and IR Photometric Data}\\
\\
\hline
\hline\\
Regions and& $(B-V)$& $(V-I)$& $V$  & $(U-B)$& $(I-K)$& $(H-K)$&$(K'-L')$& 
(A$_{V}$)$^{c}$ \\

[offset$^{b}$ ($''$)]   
       &        &        &      &        &        &        &         &       \\
\hline

Reg 1 [00-00-]
       & 0.93   & 1.55   & 14.13& ---    & 3.09   & 0.48   & 0.95 & $\geq$2.4 \\
       &        &        &      &        &        &        &      &$\geq$[7.8]\\
Reg 2 [01N05E]
       & 0.66   & 1.26   & 14.28& ---    &  3.26  & 0.61   &  --- &$\geq$1.7 \\
Reg 3 [05S00-]
       & 1.98   & 2.10   & 16.37& ---    & 2.01   & 0.20   & 1.93 &$\geq$5.3  \\
       &        &        &      &        &        &        &     &$\geq$[10.7]\\
Reg 4 [03N08E]
       & 0.63   & 1.24   & 14.51& ---    & ---    & ---    &  --- &$\geq$1.7  \\
Reg 5 [01N17E]
       & 0.71  &  1.31   & 16.42& ---    & ---    & ---    &  --- &$\geq$2.0  \\
Reg 6 [03N06W]
       & 0.94  &  1.70   & 15.37& ---    & ---    & ---    &  --- &$\geq$2.3  \\
Reg 7a[07S12W]
       & 0.98  &  1.64   & 15.51&  0.38  & ---    & ---    &  --- &$\geq$3.0  \\
Reg 7b[05S19W]
       & 1.14  &  0.91   & 15.51& ---    & ---    & ---    &  --- &$\geq$3.0  \\
Reg 8 [04N04E]
       & 0.64   & 1.27   & 14.57& ---    & ---    & ---    &  ---    & ---  \\
Reg 9 [04N01E]
       & 0.77   & 1.34   & 14.43& ---    & ---    & ---    &  ---    & ---  \\
Reg 10[07N05W]
       & 0.83   & 1.33   & 15.03& ---    & ---    & ---    &  ---    & ---  \\
Reg 11[10N03W]
       & 0.76   & 1.16   & 15.45& ---    & ---    & ---    &  ---    & ---  \\
Reg 12 [16N01E]
       & 0.77   &  1.33  & 16.37& ---    & ---    & ---    &  ---    & ---  \\
Reg 13 [12N00-]
       & 0.59   &  1.41  & 15.86& ---    & ---    & ---    &  ---    & ---  \\
Reg 14 [13N7E]
       & 0.84   &  1.33  & 16.28& ---    & ---    & ---    &  ---    & ---  \\
\hline

\end{tabular}
}$$

\newpage

$${
\begin{tabular}{lcccccccc}
\multicolumn{9}{c}{\bf Table 3: Continuation}\\
\\
\hline
Reg 15 [13N13W]
       & 0.74   &  1.36  & 16.45& ---    & ---    & ---    &  ---    & ---  \\
Reg 16 [06S06E]
       & 0.58   &  1.86  & 14.96& ---    & ---    & ---    &  ---    & ---  \\
Reg 17 [13S10E]
       & 0.85   &  1.48  & 17.26& ---    & ---    & ---    &  ---    & ---  \\
Reg 18 [12N14W]
       & 0.98   &  1.34  & 16.97& ---    & ---    & ---    &  ---    & ---  \\
Reg 19 [18S16W]
       & 0.69   &  1.28  & 17.26& ---    & ---    & ---    &  ---    & ---  \\

E-Tail 1 [46N11E]
       &  0.86  &  1.44  & 18.81& ---    & ---    & ---    &  ---    & ---  \\
E-Tail 2 [55N23E]
       &  0.71  &  1.23  & 18.77& ---    & ---    & ---    &  ---    & ---  \\
E-Tail 3 [57N65E]
       &  0.88  &  1.35  & 19.04& ---    & ---    & ---    &  ---    & ---  \\
E-Tail 4 [48N111E]
       &  0.85  & 1.11   & 19.51& ---    & ---    & ---    &  ---    & ---  \\

W-Tail 1 [09N83W]
       &  0.58  & 0.98   & 19.96& ---    & ---    & ---    &  ---    & ---  \\
W-Tail 2 [71N159W]
       &  0.51  & 1.21   & 19.98& ---    & ---    & ---    &  ---    & ---  \\
W-Tail 3 [73N159W]
       & 0.62  &  1.01   & 19.85& ---    & ---    & ---    &  ---    & ---  \\
W-Tail 4 [129N198W]
       & 0.64   & 0.35   & 20.39& ---    & ---    & ---    &  ---    & ---  \\

SW Loop [71S41W]
       & 1.05   & 1.59   & 20.03& ---    & ---    & ---    &  ---    & ---  \\
\hline

\end{tabular}
}$$

\noindent
$^{a}$: References: (B-V), (V-I) and V this work (aperture radii of 
 $\sim$ 3$''$); and (U-V), (I-K), (K'-L'), A$_{V}$ from Kotilainen et al. 
(1995) and Zenner \& Lenzen (1993);

\noindent
$^{b}$: Offset from northern nucleus.

\noindent
$^{c}$: values of internal absorption obtained from  
optical lines emission (from Table 4) and from IR lines/continuum emission 
(from Kotilainen et al. 1995, values in ``[").
These are lower  limit according the ISO results (Rigopoulou et al. 1996).
 
\newpage

$${
\begin{tabular}{lccccccc}
\multicolumn{8}{c}{\bf Table 4: Fluxes of Emission Lines$^{a}$}\\
\hline
\hline

Lines/Ratios/EqW       &      &      &       &Fluxes$^{b}$& &      &          \\
                       & Reg 1&Reg2+& Reg 4 & Reg 5& Reg 6 &Reg7ab&Reg 
1+2$^{c}$   \\

[offset$^{d}$ ($''$)]  
                       &[00-00-]&[01N05E]&[03N8E]&[02N18E]&[03N06W]&[07S12W]& \\
\hline

[O {\sc ii}]${\lambda3727}$   
                       & ---  & ---  & ---   &  --- & ---  & ---  &  72.5    \\
H$\delta\lambda4101$$^{e}$   
                       &(4.0) &(3.0) & ---   &  --- & ---  & ---  &   8.6   \\
H$\gamma\lambda4340$   &  7.0 &  8.3 &  2.2  &  1.4 & 2.5  & ---  &  14.6   \\

[O {\sc iii}]${\lambda4363}$  
                       & ---  & ---  & ---   &  --- & ---  & ---  &  (3.0) \\

H$\beta\lambda4861$    & 21.0 & 24.0 &  7.3  &  4.6 &  8.6 &(11.2)&  43.5    \\

[O {\sc iii}]
      ${\lambda5007}$  &  7.0 & 10.6 &  3.9  &  2.3 &  4.0 & (4.8)&  19.5   \\

[Na {\sc i}] D
   ${\lambda5890}$     &  --- &  3.2 &  1.0  &  --- &  1.4 & (0.7)&   3.5   \\

[O {\sc i}]
    ${\lambda6300}$    &  3.5 &  3.0 &  1.3  &  1.2 &  2.0 & (4.0)&   6.3    \\

H$\alpha\lambda6563$   &112.0 &115.0 & 36.1  & 24.1 & 52.0 & 80.8 & 219.0    \\

[N {\sc ii}]
     ${\lambda6583}$   & 65.0 & 47.7 & 13.0  &  9.0 & 21.8 & 34.4 &  85.0    \\

[S {\sc ii}]
     ${\lambda6717}$   & 13.0 & 14.3 &  5.4  &  3.0 &  9.1 & 18.8 &  30.0   \\

[S {\sc ii}]
     ${\lambda6731}$   & 13.6 & 13.0 &  4.4  &  2.3 &  7.3 & 14.0 &  30.4    \\

[S {\sc iii}]
     ${\lambda9069}$   & ---  & ---  &  ---  &  --- &  --- & ---  &  23.0   \\

[S {\sc iii}]
     ${\lambda9532}$   & ---  & ---  &  ---  &  --- &  --- & ---  &  61.0   \\

                       &      &      &       &      &      &      &         \\

H$\alpha$/H$\beta$     &  5.4 &  4.7 &  5.0  &  5.2 &  6.1 & (7.2)&  5.0    \\
E(B-V)$_{I}$           &  0.6 &  0.5 &  0.5  &  0.6 &  0.7 &  0.9 &  0.5     \\
Lum H$\alpha$ $^{f}$   &  1.8 &  2.0 &  0.6  &  0.4 &  0.9 &  1.3 &  3.6    \\
                       & [7.3]& [6.2]& [1.9] & [1.6]& [4.3]&  --- &[11.4]   \\
EqW [O {\sc ii}] (\AA) & ---  & ---  &  ---  &  --- &  --- & ---  &  44.0   \\
EqW H$\beta$ (\AA)     &  23  &  32  &  45   &  70  &  45  & (28) &  31.5   \\
EqW [O {\sc iii}] (\AA)&   8  &  14  &  30   &  38  &  20  &  (7) &  17.0   \\
EqW H$\alpha$ (\AA)    & 142  & 235  & 280   & 370  & 210  & 100  & 295.0   \\
EqW [N {\sc ii}] (\AA) &  77  &  97  & 100   & 145  &  85  &  43  & 135.0   \\
\hline

\end{tabular}
}$$

\noindent
$^{a}$: Observations from CASLEO and CTIO (see the text). 

\noindent
$^{b}$: The fluxes are given in units of 10$^{-14}$ erg cm$^{-2}$ s$^{-1}$ 
and corrected for atmospheric extinction, galactic reddening (E(B-V)=0.13) 
and redshift. The region 2+ is the region 2 plus the superposed contribution of
the border of region 4 (see the text).

\noindent
$^{c}$: Observations from CTIO.

\noindent
$^{d}$: Offset from northern nucleus.

\noindent
$^{e}$: The values between parenthesis are data with low S/N.

\noindent
$^{f}$: The luminosities are given in units of 10$^{41}$  erg 
s$^{-1}$. The values between ``[" are data corrected by internal reddening.

\newpage

$${
\begin{tabular}{lcccccc}
\multicolumn{7}{c}{\bf Table 5: Emission Line Ratios}\\
\\
\hline
\hline
          &         &         &         &         &         &        \\
Regions  & lg[O{\sc iii}]/H$\beta$$^{a}$ & lg[O{\sc i}]/H$\alpha$$^{a}$ 
& lg[N{\sc ii}]/H$\alpha$$^{a}$  & lg[S{\sc ii}s]/H$\alpha$$^{a}$ &
lg[O{\sc ii}+O{\sc iii}]/H$\beta$$^{a}$ & S{\sc ii}/S{\sc ii}$^{a}$ \\
          &         &         &         &         &         &        \\
\hline

 Reg 1$^{b}$&  -0.47&  -1.50  &  -0.27  &  -0.60  & ---     &  0.96  \\
          & [-0.50] & [-1.47] & [-0.28] & [-0.63] & ---     &        \\
 Reg 2+   &  -0.38  &  -1.60  &  -0.40  &  -0.60  & ---     &  1.10  \\
          & [-0.41] & [-1.57] & [-0.40] & [-0.62] & ---     &        \\
 Reg 4    &  -0.27  &  -1.44  &  -0.44  &  -0.57  & ---     &  1.23  \\
          & [-0.30] & [-1.42] & [-0.45] & [-0.58] & ---     &        \\
 Reg 5    &  -0.30  &  -1.30  &  -0.43  &  -0.66  & ---     &  1.30  \\
          & [-0.33] & [-1.27] & [-0.43] & [-0.68] & ---     &        \\
 Reg 6    &  -0.33  &  -1.42  &  -0.38  &  -0.50  & ---     &  1.24  \\
          & [-0.36] & [-1.38] & [-0.38] & [-0.53] & ---     &        \\
 Reg 7ab$^{c}$ 
          & (-0.38) & (-1.30) &  -0.37  &  -0.39  & ---     &  1.34  \\
          &[(-0.41)]&[(-1.26)]& [-0.38] & [-0.42] & ---     &        \\
 Reg1+2   &  -0.36  &  -1.54  &  -0.41  &  -0.56  &   0.33  &  0.98  \\ 
          & [-0.37] & [-1.52] & [-0.42] & [-0.56] &  [0.44] &        \\
\hline
\end{tabular}
}$$

\noindent
$^{a}$: [O {\sc iii}]${\lambda5007}$; [O {\sc i}]${\lambda6300}$; 
[N {\sc ii}]${\lambda6583}$; [S {\sc ii}s]${\lambda\lambda6716+6731}$;
[O {\sc ii}]${\lambda3727}$; [S {\sc ii}]/[S {\sc ii}]
${\lambda6716}$/${\lambda6731}$

\noindent
$^{b}$: Values between ``[" are corrected by internal reddening (the color 
excess E(B-V)$_{I}$ were obtained from Table 4).

\noindent
$^{c}$: The values between parenthesis are data with low S/N.

\newpage

$${
\begin{tabular}{lcc}
\multicolumn{3}{c}{\bf Table 6: Equivalent Width of the Absorption Lines}\\
\hline
\hline

Absorption Lines                &EqW$^{a}$ [\AA]& Comments \\
\hline
                                & Optical &          \\

Regions 1+2 (CTIO)              &         &          \\

Mg{\sc i}+ H9${\lambda3835}$    &  1.7    &          \\

Ca{\sc ii} K${\lambda3933}$     &  6.0    &          \\

Ca{\sc ii} H${\lambda3968}$     &  4.1    &          \\

Regions 1,2,6,7 (CASLEO)        &         &          \\

Na{\sc i} D${\lambda5890}$ Reg 1&  4.1    &  \\

Na{\sc i} D${\lambda5890}$ Reg 2&  2.6    & absorption+emission\\

Na{\sc i} D${\lambda5890}$ Reg 6&  5.5    & absorption+emission\\

Na{\sc i} D${\lambda5890}$ Reg 7&  5.0    & absorption+emission\\         

                                &UV \& near-IR &      \\
Regions 1+2  (IUE - CTIO)       &         &           \\

Si{\sc iv}${\lambda1400}$       & 10.0    & ${\lambda}_{obs}$= 1403.5 \AA\\

C{\sc iv}${\lambda1550}$        & 13.0    & ${\lambda}_{obs}$ = 1555.5 \AA \\

Fe{\sc iii}+Al{\sc iii}${\lambda\lambda1600-30}$       
                                &  3      & absorption blend \\

He{\sc ii}${\lambda1640}$       & (4)     & absorption+emission \\

Ca{\sc ii}${\lambda8498}$       &  1.1    &           \\

Ca{\sc ii}${\lambda8542}$       &  5.4    &           \\

Ca{\sc ii}${\lambda8662}$       &  3.4    &           \\
\hline

\end{tabular}
}$$

\noindent
$^{a}$: The spectra are corrected for atmospheric extinction, 
galactic reddening (E(B-V)=0.13) and redshift (see the text).

\newpage

\centerline{\bf FIGURE CAPTIONS}

\vspace{7mm}

\noindent
{\bf Fig. 1.--}
CCD broad-- and narrow-- band images of NGC~3256. North is up and east is 
to the left (in the HST--WFPC2 \& NICMOS images the  N-S lines are rotated 
$\sim$35$^{\circ}$ and $\sim$30$^{\circ}$ to the left, respectively).

{\bf (a)} 
B CASLEO image ($\sim$2$' \times $  2$'$, seeing 1.8$''$);

{\bf (b)} 
V NTT--EMMI image ($\sim$1.5$' \times $  1.5$'$, seeing 1.4$''$);

{\bf (c)} 
pure H$\alpha$+[N {\sc ii}]${\lambda\lambda6548,6583}$ NTT--SUSI images
($\sim$1$' \times $  1$'$, seeing 0.7$''$);

{\bf (d)} 
HST--WFPC2 F814W image ($\sim$40$'' \times $  40$''$, PSF 0.1$''$);

{\bf (e)} 
HST--WFPC2 F450W/F814W image ($\sim$10$'' \times $ 10$''$), the yellow
false-color show high values of the ratio F450W/F814W;

{\bf (f)} 
HST--NICMOS F222M image ($\sim$19$'' \times $ 19$''$).

{\bf (g-h)} 
HST--WFPC2 F450W image contour (1.5$'' \times $  1.5$''$),
for the main optical nucleus (region 1) and region 2;

\vspace{5mm}

\noindent
{\bf Fig. 2.--}
{\bf (a)}
Color map of the H$\alpha$ velocity field, for the central region 
($40''\times40''$ $\sim$ 7 kpc$\times$7 kpc; with a spatial resolution of 
$1''\times1''$, and errors of $\pm$15 km s$^{-1}$). 

{\bf (b)}
Contours of H$\alpha$ velocity field superposed  to a H$\alpha$+[N {\sc ii}] NTT
image, for the central region (with similar resolution and error  than Fig.
2a). The values of the lowest contour and interval are -120 km s$^{-1}$ and
+20 km s$^{-1}$.

{\bf (c)} 
Three rotation curves found in the {\it central region} of NGC 3256 -at PA 55$^{\circ}$,
90$^{\circ}$, 130$^{\circ}$- with spatial resolution  of 1.0$''$,
obtained first from a wide opening angle $\theta$ = 80$^{\circ}$ and then
for $\theta$ = 30$^{\circ}$ (see the text).

{\bf (d)} 
Three rotation curves obtained from the {\it nuclear region} of NGC 3256 -at PA
55$^{\circ}$, 90$^{\circ}$, 130$^{\circ}$-
with spatial resolution  of 0.5$''$, and obtained
from an opening  angle $\theta$ = 20$^{\circ}$; showing a superposed inflow
shape (see the text). 

{\bf (e)} 
Contour of H$\alpha$+[N {\sc ii}]${\lambda\lambda6548,6583}$ region of a 
spectra along PA 110$^{\circ}$ through the main optical nucleus, showing 
clearly the outflow/blue nuclear component (note that this blue component 
shows stronger [N {\sc ii}]${\lambda6583}$ emission than H$\alpha$). 

{\bf (f)} 
Outflow and Radial Velocity vs. axial distance (from the main optical nucleus)
 at PA 18$^{\circ}$.

\vspace{5mm}

\noindent
{\bf Fig. 3. (a)--(i)}
Optical spectra of six giant H{\sc ii} regions, located in the nuclear and 
central area of NGC 3256. Plus UV and near-IR spectra of the circumnuclear 
area (see the text).

\vspace{5mm}

\noindent
{\bf Fig. 4. (a)--(h)}
Emission line ratios and FWHM(H$\alpha$) vs. axial distance from the main optical 
nucleus (along $PA = 70^{\circ}$, 56$^{\circ}$, 155$^{\circ}$, 18$^{\circ}$,  
45$^{\circ}$, 155$^{\circ}$, 65$^{\circ}$, and 40$^{\circ}$).

\vspace{5mm}

\noindent
{\bf Fig. 5. (a)--(d)}
Emission line ratios and FWHM(H$\alpha$) color maps, for the central regions
and the main/strong component.
The north is up and east is to the left
(the scale of the emission line ratios include a factor of 10$^2$).
\vspace{5mm}

\noindent
{\bf Fig. 6.--}
Continuum Flux of NGC 3256, from radio to x-ray frecuency.

\end{document}